\documentclass[twoside]{article}
\usepackage[round]{natbib}
\usepackage{graphicx}
\usepackage{times}
\usepackage{amsmath}
\usepackage{amsfonts}
\usepackage{amssymb}
\usepackage{fancyhdr}
\usepackage{epstopdf}
\usepackage{float}
\usepackage{enumerate}
\usepackage[caption=false,position=top]{subfig}
\usepackage{xr}
\usepackage{url}
\usepackage[max6]{authblk}
\usepackage{color}
\usepackage{titlesec}
\usepackage{setspace}
\usepackage{abstract}
\usepackage[bottom=1.2in]{geometry}

\titleformat*{\section}{\large\bfseries}
\titleformat*{\subsection}{\large\itshape}
\titleformat*{\subsubsection}{\large\bfseries}
\titleformat*{\paragraph}{\large\bfseries}
\titleformat*{\subparagraph}{\large\bfseries}

\newcommand\abs[1]{\left|#1\right|}
\newcommand\cites[1]{\citeauthor{#1}'s\ (\citeyear{#1})}
 
\makeatletter
\renewcommand{\maketitle}{\bgroup\setlength{\parindent}{0pt}
	\begin{flushleft}
	\LARGE {\textbf{\@title}}
	
	\bigskip
		
	{\@author}
	\end{flushleft}\egroup
}
\makeatother

\fancyheadoffset{0cm}

\graphicspath{{C:/Work/Research/Pitsilou/Reviewpaper/Figures/}}	

\begin{document}

\title{An Updated Literature Review of Distance Correlation and its Applications to Time Series}
\author[1]{Dominic Edelmann}
\affil[1]{Department of Biostatistics, German Cancer Research Center, Heidelberg, Germany \\
	E-mail: dominic.edelmann@dkfz-heidelberg.de}
\author[2]{Konstantinos Fokianos}
\affil[2]{Department of Mathematics \& Statistics, Lancaster University, Lancaster, UK \\
E-mail: k.fokianos@lancaster.ac.uk \vspace{-6mm}}
\author[3]{Maria Pitsillou}
\affil[3]{Department of Mathematics \& Statistics, University of Cyprus, Nicosia, Cyprus \vspace{-6mm} \\
E-mail: pitsillou.maria@ucy.ac.cy}

\date{First Version: September 2017 \\ Revised Version: June 2018}
\maketitle
\thispagestyle{empty}

\begin{abstract}
{\bf The concept of distance covariance/correlation was introduced recently to characterize dependence among vectors of random variables. We review some statistical aspects of distance covariance/correlation function and we demonstrate its applicability to time series analysis. We will see that the auto-distance covariance/correlation function is able to identify nonlinear relationships and can be employed for testing the i.i.d.\ hypothesis. Comparisons with other measures of dependence are included.}
\end{abstract}

\noindent
{\textit{Key words:} {\small  characteristic function; distance covariance; nonlinear time series; Portmanteau test statistics; spectral density.}

\bigskip

\section{Introduction}
There has been a considerable recent interest in measuring dependence by employing the concept of \emph{distance covariance}, a new and appealing measure of dependence for random variables, introduced by \citet{Szekely}. As examples for the importance of distance correlation in applications we mention the work by \cite{martinez2014distance,richards2014interpreting} on astrophysical data; \cite{guo2014inferring,piras2014transcriptome,oliveira2015inferring} in molecular biology; \cite{zhang2014systemic} on return indices and stock market prices in the ship market; \cite{donnelly2016using} in hydrology; and \cite{kong2012using} on the association of lifestyle factors, diseases, familiar relationship and mortality.
Furthermore, there has been an increasing  number of works extending the concept of distance covariance in various scientific fields for independent data. In the first part of this paper, we review several of them. Complementary to our work, we refer to the recent paper by \cite{szekely2017energy} that gives a comprehensive overview of energy statistics and distance correlation methods.
\par
\cites{Szekely} distance covariance methodology is based on the assumption that the data are i.i.d. However, as this assumption is often violated in many practical problems, \citet{Remillard} proposed to extend the notion of distance covariance to the case of dependent data. \citet{Zhou} defined the so-called \emph{auto-distance covariance function} in a multivariate strictly stationary time series setting and \citet{FokianosPitsillouB} introduced
the \emph{matrix auto-distance covariance function}.  Compared to the i.i.d.\ case, there have been only few works regarding distance covariance in the context of time series. In the second part of this paper, we give a brief survey of these works and we review other measures of dependence, in the context of time series.
\par
The article is divided as follows. In Section 2, we give the basic definitions of distance covariance and correlation functions and present their basic properties. In Section 3, we review the main works of distance covariance by presenting its extensions and modifications along with some related tests of independence as these appeared in the statistics literature, for the case of independent data. The concept of distance covariance for the case of dependent data, and especially time series data, is presented in Section 4. Tests of serial independence based on distance covariance and other measures of dependence are reviewed in Section 5. A comparison of these tests is discussed in Section 6, by giving some simulated and real data examples and testing the i.i.d.\ hypothesis.

\section{The Distance Covariance Function}
\label{sec:dcov}

In what follows, $X$ denotes a univariate random variable and $\textbf{X}$ denotes a multivariate random vector.
\par
\citet{Szekely} introduced the distance covariance function as a new measure of dependence between random vectors, say $\textbf{X}$ and $\textbf{Y}$, of arbitrary, not necessarily equal, dimensions, $p$ and $q$ respectively. However, for the special case of $p=q=1$, this definition can be also found in \citet[Sec. 4]{Feuerverger} who considered a rank test for bivariate dependence.
The definition of distance covariance relies on the joint characteristic function of $\textbf{X}$ and $\textbf{Y}$, which is denoted by $\phi_{({\bf X},{\bf Y})}(\textbf{t},\textbf{s}) = E\Bigl[\exp\Bigl(i(\textbf{t}'\textbf{X} + \textbf{s}'\textbf{Y})\Bigr)\Bigr]$, where $(\textbf{t},\textbf{s}) \in \mathbb{R}^{p+q}$ and $i^2=-1$. The corresponding marginal characteristic function of $\textbf{X}$ and $\textbf{Y}$ are denoted by $\phi_\textbf{X}(\textbf{t}) = E\Bigl[\exp\Bigl(i\textbf{t}'\textbf{X}\Bigr)\Bigr]$ and $\phi_\textbf{Y}(\textbf{s}) = E\Bigl[\exp\Bigl(i\textbf{s}'\textbf{Y}\Bigr)\Bigr]$, respectively. The distance covariance function is defined as the nonnegative square root of a weighted $L_2$ distance between the joint and the product of the marginal characteristic functions of two random vectors, namely
\begin{eqnarray}
\label{eq:dcov}
V^2(\textbf{X},\textbf{Y}) & = & \int_{\mathbb{R}^{p+q}}{\abs{\phi_{({\bf X},{\bf Y})}(\textbf{t},\textbf{s})-\phi_\textbf{X}(\textbf{t})\phi_\textbf{Y}(\textbf{s})}^2 \omega(\textbf{t},\textbf{s}) d\textbf{t} d\textbf{s}},
\end{eqnarray}
where $\omega(\cdot,\cdot) : \mathbb{R}^{p+q} \rightarrow \mathbb{R}$ is a weight function for which the above integral exists and whose choice is discussed later on. Rescaling \eqref{eq:dcov} leads to the definition of the distance correlation function between \textbf{X} and \textbf{Y}, which is the positive square root of
\begin{equation}
\label{eq:dcor}
R^2(\textbf{X},\textbf{Y}) = \left\{
  \begin{array}{ll}
    \displaystyle{\frac{V^2(\textbf{X},\textbf{Y})}{\sqrt{V^2(\textbf{X},\textbf{X})V^2(\textbf{Y},\textbf{Y})}}}, & \hbox{$V^2(\textbf{X},\textbf{X})V^2(\textbf{Y},\textbf{Y}) > 0$;} \\
    $0$, & \mbox{otherwise.}
  \end{array}
\right.
\end{equation}
The previous display shows that the distance correlation function is a coefficient analogous to Pearson's correlation coefficient. But, unlike the classical coefficient which measures the linear relationship between \textbf{X} and \textbf{Y} and can be zero even when the variables are dependent, the distance correlation vanishes only in the case where \textbf{X} and \textbf{Y} are independent--see Fig \ref{dcor_graphs}. Further properties of \eqref{eq:dcov} and \eqref{eq:dcor} are established later. The choice of the weight function $\omega(\cdot,\cdot)$ is crucial. Choosing the non-integrable weight function of the form
\begin{eqnarray}
\label{eq:W}
\omega(\textbf{t},\textbf{s}) & = & (c_pc_q\abs{\textbf{t}}_p^{1+p}\abs{\textbf{s}}_{q}^{1+q})^{-1},
\end{eqnarray}
where $c_d  =  \pi^{(1+d)/2}/\Gamma((1+d)/2)$,
with $\Gamma(\cdot)$ denoting the Gamma function, $R^2(\textbf{X},\textbf{Y})$ given by \eqref{eq:dcor} is scale and rotation invariant. In particular, (\ref{eq:W}) is (up to a constant) the only weight function such that $V^2(\textbf{X},\textbf{Y})$ is rigid motion invariant and scale equivariant \citep{szekely2012uniqueness}; moreover the resulting {\it distance standard deviation}  defined as $V(X) := V(X,X)$ is then an axiomatic measure of spread \citep{edelmann2017distance}. An early reference discussing the particular choice of \eqref{eq:W} is that of \citet{Feuerverger}. For related work on characteristic functions, see also \citet{MeintanisIliopoulos} and \citet{Hlavkaetal}, among others. Furthermore, \citet{Bakirovetal} proposed a weight function $\omega(\cdot,\cdot)$ different than \eqref{eq:W}. However, the choice of \eqref{eq:W} yields to computational advantages.
Denote by $\abs{x}_{p}$ the Euclidean norm in $\mathbb{R}^{p}$.  The main properties of \eqref{eq:dcov} and \eqref{eq:dcor}, when \eqref{eq:W} is used, are given by the following:
\begin{enumerate}
  \item If $E\Bigl(\abs{\textbf{X}}_p+\abs{\textbf{Y}}_q\Bigr) < \infty$, then the distance correlation, $R(\textbf{X},\textbf{Y})$, satisfies $0 \leq R(\textbf{X},\textbf{Y}) \leq 1$ and $R(\textbf{X},\textbf{Y})=0$ if and only if \textbf{X} and \textbf{Y} are independent.
  \item $R(\textbf{X},\textbf{Y})$ is invariant under orthogonal transformations of the form $(\textbf{X},\textbf{Y}) \rightarrow \Bigl(\textbf{$\alpha_1$}+b_1C_1\textbf{X},\textbf{$\alpha_2$}+b_2C_2\textbf{Y}\Bigr)$, where $\textbf{$\alpha_1$}$, $\textbf{$\alpha_2$}$ are arbitrary vectors, $b_1$, $b_2$ are arbitrary nonzero numbers and $C_1$, $C_2$ are arbitrary orthogonal matrices.
  \item If $p=q=1$ and X and Y have standard normal distributions with $r = \mbox{Cov}(X,Y)$, then
   $R(X,Y) \leq \abs{r}$ and
   \begin{align}
       \label{eq:Rnormal}
       R^2(X,Y) &= \displaystyle \frac{r\ \mbox{arcsin}r + \sqrt{1-r^2}-r\ \mbox{arcsin}r/2-\sqrt{4-r^2}+1}{1+\pi/3-\sqrt{3}}.
       \end{align}
   \item When $E\bigl(\abs{\textbf{X}}^2_p+\abs{\textbf{Y}}^2_q\bigr) < \infty$ then
   \begin{equation}
   \label{eq:dcov.alt}
   V^2(\textbf{X},\textbf{Y})=E\abs{\textbf{X}-\textbf{X}^{\prime}}_p\abs{\textbf{Y}-\textbf{Y}^{\prime}}_q +E\abs{\textbf{X}-\textbf{X}^{\prime}}_pE\abs{\textbf{Y}-\textbf{Y}^{\prime\prime}}_q -2 E\abs{\textbf{X}-\textbf{X}^{\prime}}_p\abs{\textbf{Y}-\textbf{Y}^{\prime\prime}}_q.
   \end{equation}
   with $(\textbf{X}^\prime,\textbf{Y}^\prime)$ and $(\textbf{X}^{\prime\prime},\textbf{Y}^{\prime\prime})$ being independent copies of (\textbf{X},\textbf{Y}).
\end{enumerate}

\section{Estimation, Testing and Further Properties}
\subsection{Estimation}
\label{sec:dcovEst}
We consider the empirical counterparts of \eqref{eq:dcov} and \eqref{eq:dcor}, when using  the weight function \eqref{eq:W}.
The empirical distance covariance and correlation measures are functions of the double centered distance matrices of the samples. Suppose that $(\textbf{X}_i, \textbf{Y}_i)$, $i = 1, \cdots, n$, is a random sample from the joint distribution of the random vectors \textbf{X} and \textbf{Y}. Based on this sample, consider the $n \times n$ Euclidean pairwise distance matrices with elements $(a_{ij})=\Bigl(\abs{\textbf{X}_i-\textbf{X}_j}_p\Bigr)$ and $(b_{ij})=\Bigl( \abs{\textbf{Y}_i-\textbf{Y}_j}_q \Bigr)$. These matrices are double centered so that their row and column means are equal to zero. In other words, let
\begin{equation*}
A_{ij} = a_{ij} - \bar{a}_{i.} - \bar{a}_{.j} + \bar{a}_{..}, \quad B_{ij} = b_{ij} - \bar{b}_{i.} - \bar{b}_{.j} + \bar{b}_{..},
\end{equation*}
where $\bar{a}_{i.}=\Bigl(\sum_{j=1}^{n}{a_{ij}}\Bigr)/n$, $\bar{a}_{.j}=\Bigl(\sum_{i=1}^{n}{a_{ij}}\Bigr)/n$, $\bar{a}_{..}=\Bigl(\sum_{i,j=1}^{n}{a_{ij}}\Bigr)/n^2$. Similarly, we define the quantities $b_{i.}$,$b_{.j}$ and $b_{..}$. The sample distance covariance is defined by the square root of the statistic
\begin{eqnarray}
\label{eq:dcovn}
\widehat{V}^2(\textbf{X},\textbf{Y}) & = & \frac{1}{n^2}\sum_{i,j=1}^{n}{A_{ij}B_{ij}}.
\end{eqnarray}
We compute the squared sample distance correlation in terms of \eqref{eq:dcovn} by \eqref{eq:dcor} as
\begin{equation*}
\widehat{R}^2(\textbf{X},\textbf{Y}) = \left\{
  \begin{array}{ll}
    \displaystyle{\frac{\widehat{V}^2(\textbf{X},\textbf{Y})}{\sqrt{\widehat{V}^2(\textbf{X},\textbf{X})\widehat{V}^2(\textbf{Y},\textbf{Y})}}}, & \hbox{$\widehat{V}^2(\textbf{X},\textbf{X})\widehat{V}^2(\textbf{Y},\textbf{Y}) > 0$;} \\
    $0$, & \mbox{otherwise.}
  \end{array}
\right.
\end{equation*}
Furthermore, note that by multiplying out \eqref{eq:dcovn} we obtain the alternative expression
\begin{equation}
		\label{eq:sample.dcov.alt}
		\widehat{V}^2(\textbf{X},\textbf{Y}) = \frac{1}{n^2}\sum_{i,j=1}^n a_{ij} b_{ij} + \frac{1}{n^4} \sum_{i,j=1}^n a_{ij}  \sum_{i,j=1}^n b_{ij} -  \frac{2}{n^3} \sum_{i,j,k=1}^n a_{ij} \, b_{jk}.
		\end{equation}
The main properties of sample distance covariance and correlation functions are the following:
\begin{itemize}
  \item The estimators $\widehat{V}^2(\textbf{X},\textbf{Y})$ and $\widehat{R}^2(\textbf{X},\textbf{Y})$ are both strongly consistent, that is  they both converge almost surely to their population counterparts \eqref{eq:dcov} and \eqref{eq:dcor} respectively, as $n$ tends to infinity, provided
      that $E[\abs{\textbf{X}}_p] < \infty$ and $E[\abs{\textbf{Y}}_q] < \infty$.
  \item $\widehat{V}^2(\textbf{X},\textbf{Y}) \geq 0$ a.s., where the equality holds when \textbf{X} and \textbf{Y} are independent.
  \item $0 \leq \widehat{R}(\textbf{X},\textbf{Y}) \leq 1$.
  \item $\widehat{R}(\textbf{X},\textbf{$\alpha$}+b\textbf{X}C)=1$, where $\textbf{$\alpha$}$ is a vector, $b$ is a nonzero real number and $C$ is an orthogonal matrix.
\end{itemize}

	Figure \ref{dcor_graphs} illustrates a comparison between the empirical distance correlation $\widehat{R}$ and the empirical Pearson correlation $\widehat{Cor}$ for several examples of nonlinear association between two real-valued variables $X$ and $Y$, where the sample size is fixed at $n=1000$. While the Pearson correlation $\widehat{Cor}$ is very close to $0$ in these examples, signifying that there are either no or very weak linear dependencies, the distance correlation coefficient $\widehat{R}$ is substantially larger than $0$ for all cases.

\begin{figure}[!ht]
	\centering
	\includegraphics[width=0.95\textwidth]{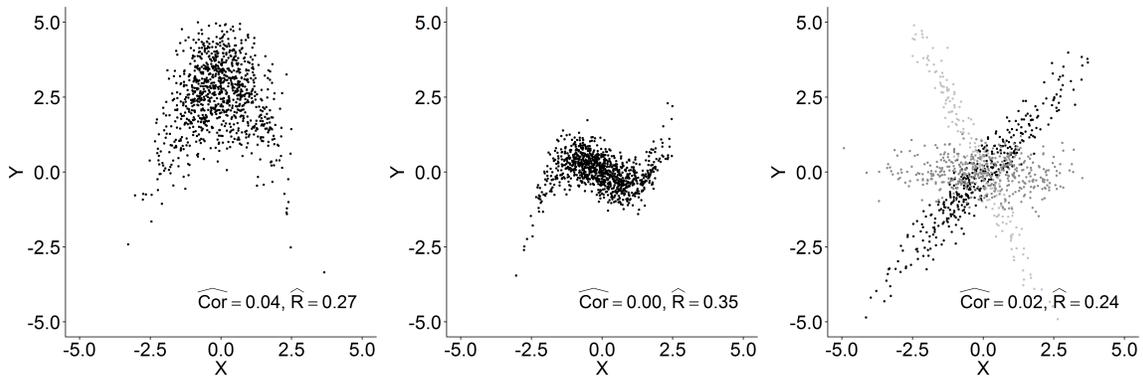}
	\caption{Comparison of empirical Pearson correlation $\widehat{Cor}$ and empirical distance correlation $\widehat{R}$ for different settings. The number of samples is fixed at $n=1000$. The left graph illustrates a quadratic dependence between $X$ and $Y$. 
		The middle graph represents a cubic relationship between $X$ and $Y$. 
		In the right graph, the relationship between $X$ and $Y$ depends on three different subgroups. 
		In the first subgroup (medium gray), $X$ and $Y$ are independent. In the second subgroup (black), there is a positive correlation between $X$ and $Y$. In the second subgroup (light gray), there is a negative correlation between $X$ and $Y$. }
	\label{dcor_graphs}
\end{figure}

It turns out that $\widehat{V}^2(\textbf{X},\textbf{Y})$ is not an unbiased estimator of $V^2(\textbf{X},\textbf{Y})$, but it holds that
\begin{eqnarray*}
E[\widehat{V}^2(\textbf{X},\textbf{Y})] & = & \frac{(n-1)(n-2)^2}{n^3} V^2(\textbf{X},\textbf{Y}) + \frac{2(n-1)^2}{n^3}\gamma - \frac{(n-1)(n-2)}{n^3}\alpha\beta,
\end{eqnarray*}
where $\alpha = E\abs{\textbf{X}-\textbf{X}'}_p$, $\beta = E\abs{\textbf{Y}-\textbf{Y}'}_q$ and $\gamma = E[\abs{\textbf{X}-\textbf{X}'}_p\abs{\textbf{Y}-\textbf{Y}'}_q]$. Here, $\textbf{X}'$ (respectively $\textbf{Y}'$) denotes an independent copy of $\textbf{X}$ (respectively, $\textbf{Y}$).
An unbiased estimator of squared distance covariance proposed by \citet{SzekelyRizzo} is given by
\begin{eqnarray*}
\label{eq:dcovn2}
\widehat{V}_U(\textbf{X},\textbf{Y}) & = & \frac{1}{n(n-3)}\sum_{i\neq j}{\widetilde{A}_{ij}\widetilde{B}_{ij}}
\end{eqnarray*}
for $n>3$, where the so-called $\mathcal{U}$-centered matrices, $\widetilde{A}_{ij}$ have the additional property that $E[\widetilde{A}_{ij}]=0$ for all $i$,$j$ and are defined by
\begin{eqnarray*}
\widetilde{A}_{ij} & = & \left\{
  \begin{array}{ll}
    a_{ij} - \frac{1}{n-2}\sum_{l=1}^{n}{a_{il}}-\frac{1}{n-2}\sum_{k=1}^{n}{a_{kj}+\frac{1}{(n-1)(n-2)}\sum_{k,l=1}^{n}{a_{kl}}}, & \hbox{$i \neq j$;} \\
    0, & \hbox{$i=j$.}
  \end{array}
\right.
\end{eqnarray*}

\par

For $p=q=1$, \citet{huo2016fast} have shown that $\widehat{V}_U(X,Y)$ is a U-statistic, which is degenerate in the case where $X$ and $Y$ are independent. For further details concerning the properties of $\widehat{V}_U(\textbf{X},\textbf{Y})$, we refer to the paper by \citet{huang2017statistically}.
By using the U-statistic representation of $\widehat{V}_U(X,Y)$, \citet{huo2016fast} show that it can be computed by an $O(n \log n)$ algorithm. This algorithm considerably speeds up the calculation of the distance correlation coefficient for large sample sizes. For higher dimensions, \citet{huang2017statistically} derive an alternative independence test based on random projections and distance correlation. Their method has
computational complexity of order  $O(K n \log n)$, where $K$ is the number of random projections.

\subsection{Asymptotic tests}
\label{sec:astests}
Following \citet{Szekely}, it can be shown that, under the null hypothesis of independence between the random vectors $\textbf{X}$ and $\textbf{Y}$,
\begin{eqnarray}
\label{eq:asympDistrVn}
n\widehat{V}^2(\textbf{X},\textbf{Y}) & \stackrel{n \to \infty}{\longrightarrow} & \sum_{j=1}^\infty \lambda_j Z_j ,
\end{eqnarray}
in distribution, where $\{Z_j\}$ are independent standard normal variables and $\{\lambda_j\}$ are eigenvalues which depend on the joint distribution of the random vectors $(\textbf{X},\textbf{Y})$, provided that $E[\abs{\textbf{X}}_p] < \infty$ and $E[\abs{\textbf{Y}}_q] < \infty$.
Moreover the quadratic form $Q:= \sum_{j=1}^\infty \lambda_j Z_j$ satisfies $E [Q] = E\abs{\textbf{X}-\textbf{X}^{\prime}}_pE\abs{\textbf{Y}-\textbf{Y}^{\prime\prime}}_q$.

On the other hand, if $\textbf{X}$ and $\textbf{Y}$ are not independent, then
\begin{eqnarray}
\label{eq:asympDistrVn.dep}
n\widehat{V}^2(\textbf{X},\textbf{Y}) &  \stackrel{n \to \infty}{\longrightarrow} & \infty
\end{eqnarray}
in probability. Results (\ref{eq:asympDistrVn}) and (\ref{eq:asympDistrVn.dep}) imply a test for independence based on the statistic $n\widehat{V}^2$. While \citet{Szekely} derive an asymptotic test with asymptotic significance level at most $\alpha$, they point out that this significance level can be quite conservative for many distributions. As an alternative, they propose a Monte-Carlo permutation test, which is now the standard method for testing independence based on the distance covariance. This permutation test is also implemented in the R package \verb"energy" by \citet{pkgenergy}. Recently, \citet{huang2017statistically} have proposed an alternative distance covariance test based on a gamma approximation of the asymptotic distribution of the test statistic under the null hypothesis.

The idea of employing \eqref{eq:dcov} for detecting independence was previously discussed by \citet{Feuerverger} who considered measures of the form of \eqref{eq:dcov}. 
To develop a nonparametric test for independence, \citet{Feuerverger} suggested to replace univariate sample points $X_i$ and $Y_i$, $i=1, \dots, n$, by approximate normal score quantities $X_i'$ and $Y_i'$, where $X_i^{\prime} = \Phi^{-1}\Biggl(\Bigl(\mbox{rank}(X_i)-3/8\Bigr)/\Bigl(n+1/4\Bigr)\Biggr)$.
Then, he proposed the test statistic
\begin{align}
\label{eq:feuerverger_stat}
\displaystyle \int_{\mathbb{R}^2}{\abs{\hat{\phi}_{({X}^\prime,{ Y}^\prime)}({t},{s})-\hat{\phi}_{X^\prime}({t})\hat{\phi}_{Y^\prime}({s})}^2\abs{t}^{-2}\abs{s}^{-2} dtds},
\end{align}
where $\hat{\phi}_{({X^\prime},{ Y^\prime})}({t},{s}) = \sum_{j=1}^{n}{e^{i(tX_j'+sY_j')}}/n$ is the empirical joint characteristic function of $(X,Y)$ and $\hat{\phi}_{X^\prime}({t}) := \hat{\phi}_{({X^\prime},{ Y^\prime})}(t,0)$ and $\hat{\phi}_{Y^\prime}({s}) := \hat{\phi}_{({X^\prime},{ Y^\prime})}(0,s)$ are the empirical marginal characteristic functions. Clearly, \eqref{eq:feuerverger_stat} is the empirical version of \eqref{eq:dcov} by using \eqref{eq:W} and is identical to \eqref{eq:asympDistrVn}; the differences are the use of scores instead of observations and the restriction to the univariate case. More on the comparison between the distance covariance and the statistics proposed by \citet{Feuerverger} can be found in \citet{Grettonetal}.
Furthermore, \citet{Feuerverger} derived conditions for the weight function $\omega(\cdot,\cdot)$ under which \eqref{eq:feuerverger_stat} can be written analogously to the right-hand side of (\ref{eq:sample.dcov.alt}), where $|\cdot - \cdot|_1$ is replaced by some function $g(\cdot,\cdot)$.
\citet{Kankainen} gave an explicit example, showing that choosing a Gaussian weight function in \eqref{eq:dcov} results in a statistic of the form (\ref{eq:sample.dcov.alt}), where $a_{ij}$ and $b_{ij}$ are replaced by $1-\exp(-(X_i-X_j)^2/(2 \sigma^2))$ and $1-\exp(-(Y_i-Y_j)^2/(2 \sigma^2))$, respectively. In a recent preprint, \citet{bottcher2017a} enlighten  the connection between the 
weight function $\omega$ and the corresponding distance $g(\cdot,\cdot)$. Furthermore, they  
give several examples including the  one given by \citet{Kankainen}.

\subsection{Generalizations and Modifications of the Distance Covariance}
\label{sec:extensionsdcov}
The $\alpha$-distance covariance function, $V^{(\alpha)}(\textbf{X},\textbf{Y})$, is a generalization of the distance covariance function \eqref{eq:dcov} for a constant $\alpha$ that lies in the interval (0,2), when \eqref{eq:W} is used. More precisely, it is the non-negative number given by the square root of \citep{Szekely}
\begin{eqnarray}
\label{eq:alpha_dist}
V^{2(\alpha)}(\textbf{X},\textbf{Y}) & = & \frac{1}{C(p,\alpha)C(q,\alpha)}\int_{\mathbb{R}^{p+q}}{\frac{\abs{\phi_{({\bf X},{\bf Y})}(\textbf{t},\textbf{s})-\phi_\textbf{X}(\textbf{t})\phi_\textbf{Y}(\textbf{s})}^2}{\abs{\textbf{t}}_p^{\alpha+p}\abs{\textbf{s}}_q^{\alpha+q}}d\textbf{t}d\textbf{s}},
\end{eqnarray}
where $C(p,\alpha)$ and $C(q,\alpha)$ are given by
\begin{eqnarray*}
C(d,\alpha) & = & \frac{2\pi^{d/2}\Gamma((1-\alpha/2))}{\alpha2^{\alpha}\Gamma((d+\alpha)/2)}.
\end{eqnarray*}
An estimator of \eqref{eq:alpha_dist} is computed analogously  by defining a distance matrix with elements
 $a_{ij} = \abs{\textbf{X}_i-\textbf{X}_j}_p^\alpha$ and $b_{ij} = \abs{\textbf{Y}_i-\textbf{Y}_j}_q^\alpha$ and working analogously as in the case of \eqref{eq:dcov}. Note that \eqref{eq:alpha_dist} extends the theory of distance covariance since \eqref{eq:dcov} is obtained as a special case when $\alpha=1$. Possible extensions of $\alpha$-distance covariance  for values of the parameter $\alpha$ which do not lie in the interval $(0,2)$ are discussed by \cite{dueck2015generalization}.

\par
On the other hand, \cite{Kosorok} suggested the weight function\[\omega(\textbf{s},\textbf{t})=c_p^{-1} c_q^{-1} (\sqrt{\textbf{s}^\prime M_1 \textbf{s}})^{-1-p} (\sqrt{\textbf{t}^\prime M_2 \textbf{t}})^{-1-q},\]  where $M_1$ and $M_2$ are positive-definite matrices. By choosing $M_1=\Sigma_\textbf{X}$ and $M_2=\Sigma_\textbf{Y}$, where $\Sigma_\textbf{X}$ and $\Sigma_\textbf{Y}$ are the covariance matrices of $\textbf{X}$ and $\textbf{Y}$, respectively we obtain the affinely invariant distance covariance and distance correlation \citep{Duecketal}, which are given, respectively, by

\begin{equation}
\begin{aligned}
\label{eq:VtildeRtilde}
\tilde{V}^2(\textbf{X},\textbf{Y}) & = V^2(\Sigma_\textbf{X}^{-1/2} \, \textbf{X}, \Sigma_\textbf{Y}^{-1/2} \, \textbf{Y}), \\
\tilde{R}^2(\textbf{X},\textbf{Y}) & = \frac{\tilde{V}^2(\textbf{X},\textbf{Y})}{\sqrt{\tilde{V}^2(\textbf{X},\textbf{X}) \, \tilde{V}^2(\textbf{Y},\textbf{Y})}} = R^2(\Sigma_\textbf{X}^{-1/2} \, \textbf{X}, \Sigma_\textbf{Y}^{-1/2} \, \textbf{Y}).
\end{aligned}
\end{equation}

$\tilde{R}$ retains all the important properties of distance correlation. In particular $0\leq \tilde{R}(\textbf{X},\textbf{Y}) \leq 1$ and $	 \tilde{R}(\textbf{X},\textbf{Y})=0$ if and only if $\textbf{X}$ and $\textbf{Y}$ are independent. Additionally it is invariant under general invertible affine transformation $\textbf{X} \rightarrow A_1\,\textbf{X}+ A_2$ and $\textbf{Y} \rightarrow B_1\,\textbf{Y}+ B_2$.

\par
An alternative way to generalize the distance covariance is based on \eqref{eq:dcov.alt} (or \eqref{eq:dcovn} if second moments do not exist). \cite{Lyons} extends distance covariance to random variables $\textbf{X}$ and $\textbf{Y}$ on general metric spaces of negative type. If $d_\mathcal{X}$ ( respectively, $d_\mathcal{Y}$) is a distance on the space where \textbf{X} is defined (respectively \textbf{Y}) and if second moments exist then the generalization is given by
\begin{align}
\label{eq:dcov.lyons}
V^2_{d_\mathcal{X},d_\mathcal{Y}}(\textbf{X},\textbf{Y}) & = E \ d_\mathcal{X}(\textbf{X},\textbf{X}^\prime)d_\mathcal{Y}(\textbf{Y},\textbf{Y}^\prime) + E \ d_\mathcal{X}(\textbf{X},\textbf{X}^\prime) \, E \ d_\mathcal{Y}(\textbf{Y},\textbf{Y}^\prime) -2 E \ d_\mathcal{X}(\textbf{X},\textbf{X}^\prime)d_\mathcal{Y}(\textbf{Y},\textbf{Y}^\prime),
\end{align}
where $\textbf{X}^\prime$, $\textbf{Y}^\prime$  are independent copies of $\textbf{X}$ and $\textbf{Y}$.
\citet{Lyons} showed that this measure characterizes independence if the underlying metric spaces are of {\it strong} negative type; in particular, this class of sets includes all separable Hilbert spaces.
\cite{Sejdinovic} extend (\ref{eq:dcov.lyons}) to semimetric spaces of negative type. They further establish the equivalence of distance covariance measures on semimetric spaces of negative type (as in (\ref{eq:dcov.lyons})) and the Hilbert-Schmidt independence criterion (HSIC), see \cite{Gretton2005,Gretton2008}. In particular, the standard distance covariance can be written as a special case of the HSIC with kernel functions given by
	$$
		k_1(\textbf{x}_1,\textbf{x}_2) = |\textbf{x}_1|_p + |\textbf{x}_2|_p - |\textbf{x}_1-\textbf{x}_2|_p, \quad \quad	 k_2(\textbf{y}_1,\textbf{y}_2) = |\textbf{y}_1|_q + |\textbf{y}_2|_q - |\textbf{y}_1-\textbf{y}_2|_q.
	$$

\par
Another extension of distance covariance was proposed by \citet{SzekelyRizzoBrownian} who considered the notion of Brownian distance covariance. Suppose that \textbf{X} and \textbf{Y} are real-valued random variables in $\mathbb{R}^p$ and $\mathbb{R}^q$ respectively, and $\textbf{W}(\textbf{s})$ and $\textbf{W}'(\textbf{t})$ are independent Brownian motions for all $\textbf{s} \in \mathbb{R}^p$ and $\textbf{t} \in \mathbb{R}^q$. Then the Brownian covariance function of \textbf{X} and \textbf{Y} is defined as the positive square root of
\begin{eqnarray}
\label{eq:process.cov}
W^2(\textbf{X},\textbf{Y}) & = & E[\textbf{X}_{\textbf{W}}\textbf{X}'_{\textbf{W}'}\textbf{Y}_{\textbf{W}}\textbf{Y}'_{\textbf{W}'}],
\end{eqnarray}
where
\begin{eqnarray*}
\textbf{X}_\textbf{W} & = & \textbf{W}(\textbf{X}) - E[\textbf{W}(\textbf{X})|\textbf{W}]
\end{eqnarray*}
and $(\textbf{W},\textbf{W}')$ are independent of $(\textbf{X},\textbf{Y},\textbf{X}',\textbf{Y}')$. The authors proved the surprising result that the Brownian covariance coincides with the squared population distance covariance \eqref{eq:dcov} \citep[Thm. 8]{SzekelyRizzoBrownian}, that is
\begin{equation*}
W^2(\textbf{X},\textbf{Y}) = 
V^2(\textbf{X},\textbf{Y}).
\end{equation*}
Further extensions of this idea to other stochastic processes, such as fractional Brownian motions and to more general Gaussian processes, like the Laplace-Gaussian process are given by \citet{SzekelyRizzoEnergy}. When $(X,Y)$ is a bivariate vector and the processes $W$ and $W^\prime$ are replaced by the identity function $id$, that is $X_{id}=X-E[X]$  and $Y_{id} = Y - E[Y]$, then we obtain the absolute value of the classical Pearson covariance.
\par
Furthermore, \citet{SzekelyRizzoEnergy} extended this idea to a general class of Gaussian processes, showing that for any stochastic processes of this form, \eqref{eq:process.cov} can be expressed similarly as the right-hand side of (\ref{eq:dcov.lyons}) (where the (semi)-metrics are replaced by conditional negative definite continuous symmetric functions generating the covariance functions of the processes). \citet{Fiedler} developed new dependence measures based on the distance correlation function for spatial stochastic processes. Moreover, a recent contribution of the appropriate definitions of distance covariance for measuring dependence in the context of stochastic processes is that of \citet{Matsuietal}.
\par
\citet{Remillard} suggested to extend the notion of distance covariance by replacing the samples $\{\textbf{X}_i\}$, $\{\textbf{Y}_i\}$ by their normalized ranks, that is $\{R_{\textbf{X},i}/n\}$ and $\{R_{\textbf{Y},i}/n\}$ respectively, where $R_{X,ij}$ is the rank of $X_{ij}$ among $X_{1j}, \dots, X_{nj}$, $j = 1, \dots, p$. He further suggested to generalize the distance covariance to measure dependence among more than two random vectors. For random vectors $\textbf{X}_1$, \dots, $\textbf{X}_d$, taking values in $\mathbb{R}^{p_1}, \dots, \mathbb{R}^{p_d}$, \citet{Belu} defined the distance covariance as the weighted difference between the $d$-dimensional joint characteristic function, \[\phi(\textbf{t}_1, \dots, \textbf{t}_d) = E\Bigl[\exp\Bigl(i(\textbf{t}'_1\textbf{X}_1 + \dots + \textbf{t}'_d\textbf{X}_d)\Bigr)\Bigr],\] and the product of the marginals \[\phi^{(j)}(\textbf{t}_j) = E\Bigl[\exp\Bigl(i(\textbf{t}'_j\textbf{X}_j)\Bigr)\Bigr], \quad j = 1, \cdots, d.\] The extended version of distance covariance (compare \eqref{eq:dcov} and \eqref{eq:alpha_dist}) is defined as the positive square root of
\begin{eqnarray*}
V^{2(\alpha)}(\textbf{X}_1, \dots, \textbf{X}_d) & = & \frac{1}{c_{p_1}...c_{p_d}}\int_{\mathbb{R}^{p_1 + \dots + p_d}}{\frac{\abs{\phi(\textbf{t}_1, \dots, \textbf{t}_d) - \phi^{(1)}(\textbf{t}_1) \cdots \phi^{(d)}(\textbf{t}_d)}^2}{\abs{\textbf{t}_1}_{p_1}^{\alpha+p_1}...\abs{\textbf{t}_d}_{p_d}^{\alpha+p_d}}},
\end{eqnarray*}
where $c_{p_j}$, $j = 1, \dots, d$ is given as before and $\alpha$ is a positive constant that lies in the interval (0,2).
The above quantity can be employed for testing independence in the case of more than two random vectors but
\citet[p. 471]{bakirov2011}  point out that this integral can diverge. 
For novel approaches of multivariate version of  distance covariance, 
see \citet{yao2018}, \citet{bottcher2017b} and \citet{ChakrabortyandZhang(2018)}.
A related line of research is based on HSIC; see \citet{Pfisteretal(2018)}.

\cite{wang2015conditional} introduced the {\it conditional distance covariance} as the square root of
	$$
	\mbox{cdCov}^2(\textbf{X},\textbf{Y}|\textbf{Z}) = \frac{1}{c_p \, c_q}\int_{\mathbb{R}^{p+q}}{\frac{\abs{\phi_{({\bf X},{\bf Y}|{\bf Z})}(\textbf{t},\textbf{s})-\phi_{{\bf X}|{\bf Z}}(\textbf{t})\phi_{{\bf Y}|{\bf Z}}(\textbf{s})}^2}{|\textbf{t}|_p^{p+1} |\textbf{s}|_q^{q+1}}  d\textbf{t} d\textbf{s}},
	$$
	where $\phi_{(\textbf{X},\textbf{Y})|\textbf{Z}}$ denotes the conditional joint characteristic function of $(\textbf{X},\textbf{Y})$ given $\textbf{Z}$ and $\phi_{\textbf{X}|\textbf{Z}}$ and $\phi_{\textbf{Y}|\textbf{Z}}$ denote the respective conditional marginal characteristic functions.
	Analogously the {\it conditional distance correlation} is defined as the square root of 	
	$$
	\mbox{cdCor}^2(\textbf{X},\textbf{Y}|\textbf{Z}) = \frac{\mbox{cdCov}^2(\textbf{X},\textbf{Y}|\textbf{Z})}{\sqrt{\mbox{cdCov}^2(\textbf{X},\textbf{X}|\textbf{Z})\mbox{cdCov}^2(\textbf{Y},\textbf{Y}|\textbf{Z})}},
	$$
	provided that $\mbox{cdCov}^2(\textbf{X},\textbf{X})|\textbf{Z}) \mbox{cdCov}^2(\textbf{Y},\textbf{Y}|\textbf{Z})>0$, and $0$, otherwise.
	It is easy to see that $\mbox{cdCov}^2(X,Y|Z)$ equals $0$ if and only if $X$ and $Y$ are independent given $Z$. In addition, \cite{wang2015conditional} derive two consistent, ready-to-use sample measures for the conditional distance covariance, which do not involve integration of the empirical characteristic functions.
	
	It is important to point out that the conditional distance correlation is not equivalent to the {\it partial distance correlation} introduced by \cite{SzekelyRizzo}, which can be defined via
	$$
	R^*(X,Y;Z) = \frac{R^2(X,Y) - R^2(X,Z) R^2(Y,Z)}{\sqrt{(1-R^4(X,Z))} \sqrt{(1-R^4(Y,Z))}},
	$$
	if $R(X,Z) \neq 1$ and $R(Y,Z) \neq 1$ and $0$ otherwise.	
	In particular, \cite{SzekelyRizzo} show that $R^*(X,Y;Z) =0$ is not equivalent to conditional independence of $X$ and $Y$ given $Z$.
	
	Another related measure is the {\it martingale difference correlation} \citep{shao2014martingale},
	which is defined for ${\bf X} \in \mathbb{R}^p$ and $Y \in \mathbb{R}$ by
	$$
	\mbox{MDC} = \frac{\mbox{MDD}(Y|\textbf{X})}{\sqrt{\mbox{var}^2(Y) V^2({\bf X},{\bf X})}}.
	$$
	Here $\mbox{MDD}(U,V)$ denotes the {\it martingale difference divergence} given by
	$$
	\mbox{MDD}(Y|\textbf{X})	= \frac{1}{c_p}\int_{\mathbb{R}^{p+q}}{\frac{\abs{E\bigl[Y \, \exp\bigl(i\textbf{t}'\textbf{X}\bigr)\bigr]-E\bigl[\exp\bigl(i\textbf{t}'\textbf{X}\bigr)\bigr]E[Y]}^2}{|\textbf{t}|_p^{p+1}}  d\textbf{t} d\textbf{s}}.
	$$
The martingale correlation satisfies the property that $\mbox{MCC}(V|U)=0$ if and only if $E[V|U]=E[V]$ almost surely.
In \cite{park2015partial}, a conditional version of the martingale difference correlation is introduced; it is defined analogously to the partial distance correlation of \cite{SzekelyRizzo}.

\subsection{Variable selection}

	Since the distance correlation coefficient captures any kind of dependence including nonlinear or non monotone dependencies, it appears natural to apply this coefficient for variable selection in nonlinear models. For this purpose, \cite{li2012feature} have developed the distance correlation sure independence screening method (DC-SIS) for variable screening. As the usual SIS \citep{fan2008sure} selects the variables with the highest Pearson correlation, DC-SIS selects the variables with the highest distance correlation. Unlike SIS, which is applicable to linear models only, DC-SIS can detect arbitrary associations of the predictors with the response. \cite{li2012feature} show the sure screening property for DC-SIS, under moderate assumptions, and show that this method accommodates linear models as a special case. Two iterative modifications for DC-SIS have been developed recently ( see \cite{yenigun2015variable,zhong2015iterative}). However there are no known theoretical properties of these methods as of yet, to the best of our knowledge. \cite{kong2015using} have derived an automatic stopping rule for DC-SIS, which determines the number of predictors to select.
	 Recently, \cite{liu2017model} have proposed a modified version of DC-SIS based on the concept of conditional distance correlation \citep{wang2015conditional}, which allows for feature screening conditional on some prespecified variables. Other notable approaches are the methods by \citet{shao2014martingale}, who propose a variable screening algorithm based on the martingale difference correlation and \citet{kong2017interaction} who derive the sure screening property for a two-stage screening procedure that applies distance correlation on transformed variables.
	
	Considering the equivalence of the generalized distance covariance and the Hilbert Schmidt Independence Criterion \citep{Sejdinovic}, more general distance correlation based variable selection methods emerge as special cases of HSIC based methods. As examples, see \cite{song2012feature}, \cite{balasubramanian2013ultrahigh} and \cite{tonde2016supervised}.

\subsection{Closed form expressions for the distance covariance of multivariate distributions}

Distance correlation possesses an astonishingly simple sample quantity but the analytical evaluation of its population quantity for known parametric distribution often represents computational problems. For distributions with finite support, it is straightforward (but cumbersome) to evaluate \eqref{eq:dcor} to obtain closed form expressions for the distance covariance. For example, for two Bernoulli-distributed random variables $X$ and $Y$ with $P\bigl((X,Y)=(0,0)\bigr) = P\bigl((X,Y)=(1,1)\bigr) = p/2$ and $P\bigl((X,Y)=(0,1)\bigr) = P\bigl((X,Y)=(1,0)\bigr) = (1-p)/2$, we have that
	$$
		R^2(X,Y)= (2p-1)^2 = \rho^2(X,Y),
	$$
where $\rho$ denotes the ordinary Pearson correlation coefficient. For distributions with non-finite support, the usual approach is to directly calculate the integral (\ref{eq:dcov}). For normally distributed $X$ and $Y$ with correlation $r$  and $EX=EY=0$, $\mbox{Var}(X)=\mbox{Var}(Y)=1$ we have already seen identity \eqref{eq:Rnormal}.
\citet{Duecketal} extended this result to the affinely invariant distance correlation (recall \eqref{eq:VtildeRtilde}) of two multivariate normal vectors $\textbf{X}$ and $\textbf{Y}$. When $Cov(\textbf{X})=I_p$ and $Cov(\textbf{Y})=I_q$ this coincides with the regular distance correlation. In this particular case, the squared affinely invariant distance covariance is given as
	\begin{eqnarray*}  \label{eq:aidcov}
	\tilde{V}^2(\textbf{X},\textbf{Y})
	= 4 \pi \frac{c_{p-1}}{c_p} \frac{ c_{q-1}}{c_q}
	\times
	\Big( {}_3F_2 \! \left(\tfrac12,-\tfrac12,-\tfrac12;\tfrac{p}{2},\tfrac{q}{2};\Lambda\right)
	- 2 \, {}_3F_2 \! \left(\tfrac12,-\tfrac12,-\tfrac12;\tfrac{p}{2},\tfrac{q}{2};\tfrac14\Lambda\right) + 1 \Big),
	\end{eqnarray*}
where $c_{p}$ has been defined immediately after \eqref{eq:W},  $\Lambda$ denotes  the squared cross covariance matrix of $\textbf{X}$ and $\textbf{Y}$ and ${}_3F_2$ denotes a generalized hypergeometric function of matrix argument (see \citet{Duecketal} for the details).
\citet{DeuckLancaster} outlined a general approach for calculating the distance correlation for the so-called Lancaster class of distributions. The authors illustrate this result by calculating the distance correlation for the bivariate gamma, the bivariate Poisson and the bivariate negative binomial since
all of those distributions belong to the  Lancaster class. Moreover, they derive an alternative expression for the affinely invariant distance correlation of the multivariate normal distribution   and derive the distance correlation for generalizations of the bivariate normal.

\subsection{The distance correlation coefficient in high dimensions}

As stated in subsection \ref{sec:astests}, distance covariance provides a test for independence for random vectors $\textbf{X}$ and $\textbf{Y}$ of arbitrary dimension $p$ and $q$, respectively. However, practitioners often use distance correlation rather as an an ad-hoc measure for independence than applying a formal test. While this appears to work well for univariate random variables, some authors have reported that the direct interpretation of distance correlation is questionable in high dimensions. \citet{Duecketal} showed that for fixed $q$ and standard multivariate normal random vectors $(\textbf{X},\textbf{Y}) \in \mathbb{R}^{p+q}$
\begin{equation} \label{eq:duecklim}
\lim_{p \to \infty}	  {R}(\textbf{X},\textbf{Y})  = 0,
\end{equation}
irrespectively of the dependence structure between $\textbf{X}$ and $\textbf{Y}$.
On the other hand, \citet{Szekelyttest} proved that for fixed $n$
	\begin{equation} \label{eq:szekelylim}
	  \lim_{p,q \to \infty}	  \hat{R}(\textbf{X},\textbf{Y})  = 1,
	\end{equation}
provided that the coordinates of $\textbf{X}$ and $\textbf{Y}$ are i.i.d.\ and the second moments of $\textbf{X}$ and $\textbf{Y}$ exist.
Comparison of  (\ref{eq:duecklim}) and (\ref{eq:szekelylim}) reveals that there are actually two different effects, which complicate the interpretation of distance correlation in high dimension. Equation (\ref{eq:duecklim})  shows  that the {\it population} distance correlation is close to $0$ when comparing a low-dimensional and a high-dimensional vector. Equation (\ref{eq:szekelylim} shows that the {\it sample} distance correlation is close to $1$ when comparing two high-dimensional vectors. These two problems may occur simultaneously, when comparing two high-dimensional vectors with substantially distinct dimensions. In particular, (\ref{eq:duecklim}) shows that we can find sequences $(p_k)_{k \in \mathbb{N}},(q_k)_{k \in \mathbb{N}}$, such that $p_k \to \infty$, $q_k \to \infty$ and
  \begin{equation*} \label{eq:szekelylim2}
  \lim_{k \to \infty}	  {R}(\textbf{X},\textbf{Y})  = 0, \quad \quad \lim_{k \to \infty} \hat{R}(\textbf{X},\textbf{Y})  =1,	
  \end{equation*}
where for any $k$, $\textbf{X}$ and $\textbf{Y}$ are $p_k$ and $q_k$-dimensional standard normal vectors and $n$ is fixed.
When $p=q$, the population distance correlation ${R}(\textbf{X},\textbf{Y})$ can attain both the values $0$ ($\textbf{X}$ and $\textbf{Y}$ independent) and $1$ ($\textbf{Y}=\alpha + b \ C \, \textbf{X}$ with $\alpha \in \mathbb{R}^p,\, b \in \mathbb{R},\, C \text{ orthogonal}$). In view of (\ref{eq:szekelylim}) we can apply an unbiased estimator for the distance covariance as described in \citet{Szekelyttest}. This has the further advantage that transformation of this estimator converges to a Student-$t$ distribution with $n(n-3)/2$ degrees of freedom, which enables a distance correlation $t$-test for independence in high dimensions. When $p$ and $q$ differ (but are both high-dimensional), the $t$-test may still be applied, however a direct interpretability is critical (recall equation \eqref{eq:duecklim}). \\


\section{Auto - Distance Covariance Function}
\label{sec:dcovTS}

\subsection{Basic definitions}

The distance covariance methodology discussed in Section \ref{sec:dcov} is based on the assumption that the observations are i.i.d. However, in many practical problems this assumption is violated. \citet{Remillard} proposed an extension of the distance covariance methodology to non-i.i.d.\ observations, especially time series data, for measuring serial dependence. In this section, we review and compare a number of related dependence measures that have appeared in the time series literature. Throughout the remainder of the paper, we denote by bold $\{\textbf{X}_t\}$ a $d$-dimensional stationary time series process, with components $\{X_{t;i}\}_{i=1}^d$. Define further the joint characteristic function of $\textbf{X}_{t}$ and $\textbf{X}_{t+j}$ by $\phi_{j}(\textbf{u},\textbf{v}) = E\Bigl[\exp\Bigl(i(\textbf{u}'\textbf{X}_t + \textbf{v}'\textbf{X}_{t+j})\Bigr)\Bigr]$ and the marginal characteristic function of $\textbf{X}_t$ by $\phi(\textbf{u}) = E\Bigl[\exp\Bigl(i\textbf{u}'\textbf{X}_t\Bigr)\Bigr]$, with $(\textbf{u},\textbf{v}) \in \mathbb{R}^{2d}$, $i^2=-1$ and $j \in \mathbb{Z}$. The above definitions include the case of univariate time series; that is $d=1$ and $(u,v) \in \mathbb{R}^2$.

\par
\citet{Zhou} introduced the definition of distance covariance in time series literature. In particular, extending \cites{Szekely} definition, \citet{Zhou} defined the auto-distance covariance function (ADCV) as the positive square root of \eqref{eq:dcov}, between the joint and the product of the marginal characteristic functions of $\textbf{X}_t$ and $\textbf{X}_{t+j}$. However,
\citet{HongB} had already defined a similar measure of dependence that captures all types of pairwise dependencies. In what follows, we explain how this work is related to \cites{Szekely} work for the definition of distance covariance in time series.

\par
For a strictly stationary $\alpha$-mixing univariate time series $\{X_t\}$, \citet{HongB} defined a new measure of dependence between the joint characteristic function of $X_t$ and its lagged observation $X_{t+j}$ and the product of their marginals, namely
\begin{align}
\label{eq:sigma}
\sigma_j(u,v) & = \mbox{Cov}\bigl(e^{iuX_t},e^{ivX_{t+j}}\bigr) = \phi_{j}(u,v) - \phi(u)\phi(v),
\end{align}
where $j \in \mathbb{Z}$. Given that the joint characteristic function factorizes into the product of its marginals under independence of $X_t$  and $X_{t+j}$, $\sigma_{j}(u,v)$ equals 0 if and only if $X_t$ and $X_{t+j}$ are independent. Thus, compared to the classical autocorrelation function (ACF),
\eqref{eq:sigma} can capture all pairwise dependencies including those with zero autocorrelation.
Note that various measures of dependence can be based on \eqref{eq:sigma} (recall \eqref{eq:dcov}) by considering
symmetric weighted integrals of the form (see also \citet{Davisetal})
\begin{equation*}
\label{eq:intSigma}
\int{\abs{\sigma_j(u,v)}^2 \omega(u,v)dudv},
\end{equation*}
for a suitable function $\omega(\cdot,\cdot)$. For instance, consider $\omega(u,v)=\phi(u)\phi(v)$ where $\phi(\cdot)$ is the density of the standard normal.
Motivated by the work of \citet{Szekely}, \citet{Zhou} also defined the so-called auto-distance covariance function of $\{X_t\}$ as the positive square root of
\begin{eqnarray}
\label{eq:Vpop}
V_X^2(j) & = & \frac{1}{\pi^2}\int_{\mathbb{R}^2}\frac{\abs{\sigma_j(u,v)} ^2}{\abs{u}^2\abs{v}^2} du dv, \quad j \in \mathbb{Z}.
\end{eqnarray}
From the above definition, we can observe that $V_X^2(j)$ equals 0 if and only if $X_t$ and $X_{t+j}$ are independent.
Rescaling \eqref{eq:Vpop}, we define the auto-distance correlation function (ADCF) as the positive square root of
\begin{equation}
\label{eq:R}
R_X^2(j) = \left\{
  \begin{array}{ll}
    \displaystyle{\frac{V_X^2(j)}{V_X^2(0)}}, & \hbox{$V_X^2(0) \neq 0$;} \\
    $0$, & \hbox{$V_X^2(0) = 0$.}
  \end{array}
\right.
\end{equation}
Clearly, $R_{X}^2(j)$ equals zero if and only if $X_t$ and $X_{t+j}$ are independent. Thus, the ADCF is suitable for exploring possible nonlinear dependence structures in time series that are not detected by the ACF.
This is better understood by examining Figure \ref{fig:NMA2univ}.
The plot presents the ACF and ADCF of a second order nonlinear moving average process (NMA(2)) defined by
\begin{eqnarray}
\label{eq:NMA2model}
X_t & = & \epsilon_t \epsilon_{t-1} \epsilon_{t-2},
\end{eqnarray}
where $\{\epsilon_t\}$ is a sequence of i.i.d.\ standard normal random variables. It is well known that the process $\{X_t\}$ consists of uncorrelated but two-dependent random variables. The white noise property is clearly discovered by the ACF, whereas the dependence structure is only detected by the ADCF. We note that the shown critical values of the ADCF plots (second column) are the pairwise $95\%$ critical values that correspond to the pairwise independence test. They are computed via the subsampling approach suggested by \citet[Section 5.1]{Zhou}, where the choice of the block size is based on the minimum volatility method proposed by \citet[Section 9.4.2]{PolitisBook}. Moreover, the $95\%$ simultaneous critical values, for testing whether the data form an iid sequence, are also shown in the sample ADCF plots (last column). They are computed via the wild bootstrap method \citep{DehlingMikosch,Shao2010,Leuchtb} following the algorithm presented in \citet{FokianosPitsillouB}.

\begin{figure}
  \includegraphics[width=17cm,height=7cm]{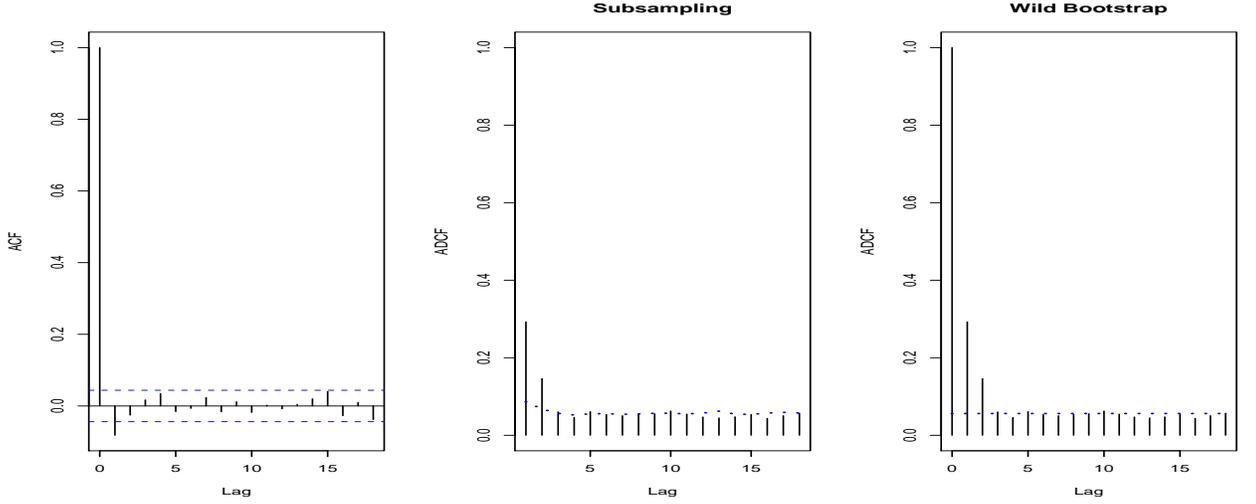}\\
  \caption{\small{Comparison of the sample ADCF and sample ACF. The data are generated by the NMA(2) model given by \eqref{eq:NMA2model}. Results are based on sample size $n=2000$.}}
  \label{fig:NMA2univ}
\end{figure}

\subsection{Estimation of ADCV}
\label{sec:ADCVest}
The sample ADCV, denoted by $\widehat{V}_X(\cdot)$, is defined analogously as in Section \ref{sec:dcovEst}, with the second sample replaced by the lagged observation of the time series, that is by letting $Y_t=X_{t+j}$ for $j \geq 0$. In particular, by defining the $(n-j) \times (n-j)$ pairwise Euclidean distance matrices by $(a_{rl}) = \abs{X_{r}-X_{l}}$ and $(b_{rl}) = \abs{Y_{r}-Y_{l}}$, the sample ADCV is given by
\begin{eqnarray}
\label{eq:sampleV}
\widehat{V}_X^2(j) & = & \frac{1}{(n-j)^2}\sum_{r,l=1}^{n-j}{A_{rl}B_{rl}}.
\end{eqnarray}
where $A_{rl} = a_{rl} - \bar{a}_{r.} - \bar{a}_{.l} + \bar{a}_{..}$ and $B_{rl} = b_{rl} - \bar{b}_{r.} - \bar{b}_{.l} + \bar{b}_{..}$,
with $\bar{a}_{r.}=\Bigl(\sum_{l=1}^{n-j}{a_{rl}}\Bigr)/(n-j)$, $\bar{a}_{.l}=\Bigl(\sum_{r=1}^{n-j}{a_{rl}}\Bigr)/(n-j)$ and $\bar{a}_{..}=\Bigl(\sum_{r,l=1}^{n-j}{a_{rl}}\Bigr)/(n-j)^2$. Similarly, we define the quantities $b_{r.}$,$b_{.l}$ and $b_{..}$. If $j<0$, then $\widehat{V}^2_X(j) = \widehat{V}^2_X(-j)$. Replacing \eqref{eq:sampleV} into \eqref{eq:R}, the sample ADCF is the nonnegative number defined by the square root of

\begin{equation*}
\widehat{R}_X^2(j) = \left\{
  \begin{array}{ll}
    \displaystyle{\frac{\widehat{V}_X^2(j)}{\widehat{V}_X^2(0)}}, & \hbox{$\widehat{V}_X^2(0) \neq 0$;} \\
    $0$, & \hbox{$\widehat{V}_X^2(0) = 0$.}
  \end{array}
\right.
\end{equation*}

The main properties of ADCV and ADCF and their empirical counterparts are summarized below:
\begin{enumerate}
  \item Under the minimal conditions of a strictly stationary $\alpha$-mixing process and existence of  first moment, \citet{FokianosPitsillou} obtained the strong consistency of $\widehat{V}_X^2(\cdot)$ to its population counterpart for a fixed lag $j$. \citet{Zhou} obtained the weak consistency of $\widehat{V}_X(\cdot)$ requiring $(1+r_0)$;th  moment of the process for some $r_0 > 0$ by assuming short range dependence condition. \citet{Davisetal} also proved strong consistency for ADCV, under various choices of finite and infinite weight functions $\omega(\cdot,\cdot)$.
      In addition, \citet{FokianosPitsillouB} show strong  consistency of  $\widehat{V}_X^2(\cdot)$  by assuming existence of second order moments of $X_{t}$.  The proof is simpler when compared to other proofs but at the cost of assuming $\mbox{E}|X_{t}|^{2} < \infty$.
  \item $\widehat{V}_X^2(j) \geq 0$ a.s., for $j \in \mathbb{Z}$ where the equality holds under independence.
  \item If $X_t$ has finite first moments, then $0 \leq R_X(j) \leq 1$. $R_X(j) = 0$ if and only if $X_t$ and $X_{t+j}$ are independent and $R_X(j) = 1$ if $X_t$ and $X_{t+j}$ are linearly related by an orthogonal matrix.
  \item If $\{X_t\}$ is a univariate Gaussian zero-mean stationary time series process with $\mbox{Var}(X_t)=1$ and $\rho(j)=\mbox{Cov}(X_t,X_{t+j})$ then $R_X(j) \leq \abs{\rho(j)}$ and
      \begin{align*}
      R_X^2(j) & =  \displaystyle \frac{\rho(j) \ \mbox{arcsin}\rho(j) + \sqrt{1-\rho^2(j)}-\rho(j) \ \mbox{arcsin}\rho(j)/2-\sqrt{4-\rho^2(j)}+1}{1+\pi/3-\sqrt{3}}, \quad j \in \mathbb{Z}.
      \end{align*}
\end{enumerate}

\subsection{Extensions of ADCV}
\label{sec:extensionsADCV}
\citet{Zhou} defined the distance correlation coefficient to explore temporal nonlinear dependence structure in multivariate time series. However, his approach is missing the interrelationships between the various time series components.
For a strictly stationary $d$-dimensional time series $\{\textbf{X}_t\}$, \citet{Zhou} defined
\begin{eqnarray}
\label{eq:V}
V_\textbf{X}^2(j) & = & \frac{1}{c_d^2}\int_{\mathbb{R}^d}\int_{\mathbb{R}^d}\frac{\abs{\sigma_j(\textbf{u},\textbf{v})} ^2}{\abs{\textbf{u}}^{d+1}\abs{\textbf{v}}^{d+1}} d\textbf{u} d\textbf{v}, \quad j \in \mathbb{Z}.
\end{eqnarray}
When testing for independence for multivariate time series, a natural extension consists of defining the distance correlation appropriately for examining whether there exists some inherent nonlinear interdependence between the component series. \citet{FokianosPitsillouB} investigated this problem by defining the \emph{pairwise auto-distance covariance and correlation matrices}, and extended the theory described in Section \ref{sec:testsADCV} to multivariate processes. In particular, define an analogous dependence measure to \eqref{eq:sigma} by
\begin{eqnarray*}
\label{eq:sigmarmk}
\sigma_{j}^{(r,m)}(u,v) & = & \phi_{j}^{(r,m)}(u,v) - \phi^{(r)}(u)\phi^{(m)}(v),  \quad  j \in \mathbb{Z},
\end{eqnarray*}
where $\phi_{j}^{(r,m)}(u,v)$, $\phi^{(r)}(u)$ and $\phi^{(m)}(v)$ are the joint characteristic function and the marginal characteristic functions of $X_{t;r}$ and $X_{t+j;m}$ respectively, for $r,m=1, \dots, d$. Then the pairwise auto-distance covariance function between $X_{t;r}$ and $X_{t+j;m}$ as the positive square root of
\begin{eqnarray}
\label{eq:Vrm}
V_{rm}^2(j) & = & \int_{\mathbb{R}^2}{\abs{\sigma_j^{(r,m)}(u,v)} ^2 \omega(u,v) du dv}, \quad j \in \mathbb{Z},
\end{eqnarray}
where $\omega(\cdot,\cdot)$ is given in \eqref{eq:W}. The auto-distance covariance matrix of $\{\textbf{X}_t\}$ at a specific lag $j$, denoted by $V(j)$, is the $d \times d$ matrix \[V(j) = \Biggl[V_{rm}(j)\Biggr]_{r,m=1}^d, \quad j \in \mathbb{Z}.\]
The pairwise auto-distance correlation function between $X_{t;r}$ and $X_{t+j;m}$, say $R_{rm}(j)$, is then defined by rescaling \eqref{eq:Vrm}
\begin{eqnarray*}
\label{eq:Rrm}
R^2_{rm}(j) & = & \frac{V_{rm}^2(j)}{\sqrt{V^2_{rr}(0)}\sqrt{V^2_{mm}(0)}},
\end{eqnarray*}
for $V_{rr}(0)V_{mm}(0) \neq 0$ and zero otherwise. The auto-distance correlation matrix at lag $j$, denoted by $R(j)$, is the $d \times d$ matrix
\[R(j) = \Biggl[R_{rm}(j)\Biggr]_{r,m=1}^d, \quad j \in \mathbb{Z}.\]
The pairwise distance correlation coefficient  can be employed for
exploring complex nonlinear interdependence structures that can not be discovered by the classical cross-correlation coefficient.
\par
The sample quantity of \eqref{eq:Vrm} is obtained analogously as in Section \ref{sec:ADCVest}. In particular, for $j \geq 0$ consider a sample $(X,Y) = \{(X_{t;r},X_{t+j;m})\}$ for $r,m =1, 2, \dots, d$. Define the Euclidean distance matrices by $(a_{ts}^r)=\abs{X_t-X_s}$ and $(b_{ts}^m)=\abs{Y_t-Y_s}$ and the double centered distance matrices by
\begin{equation*}
A_{ts}^r = a_{ts}^r - \bar{a}_{t.}^r - \bar{a}_{.s}^r + \bar{a}_{..}^r, \quad B_{ts}^m = b_{ts}^m - \bar{b}_{t.}^m - \bar{b}_{.s}^m + \bar{b}_{..}^m,
\end{equation*}
where the quantities in the right hand side are defined analogously as those defined in Section \ref{sec:ADCVest}. Thus, the sample counterpart of $V_{rm}(\cdot)$ is the positive square root of
\begin{eqnarray*}
\label{eq:Vrmest2}
\widehat{V}_{rm}^2(j) & = & \frac{1}{(n-j)^2}\sum_{t,s=1}^{n-j}{A_{ts}^{r} B_{ts}^{m}}.
\end{eqnarray*}
If $j < 0$, set $\widehat{V}_{rm}^2(j) = \widehat{V}_{mr}^2(-j)$. The sample distance covariance matrix is then given by \[\widehat{V}(j) = \Biggl[\widehat{V}_{rm}(j)\Biggr]_{r,m=1}^d, \quad j \in \mathbb{Z}.\]
The main properties of $V_{rm}(\cdot)$ and $R_{rm}(\cdot)$ and their sample counterparts are outlined below:
\begin{itemize}
  \item For all $j \in \mathbb{Z}$, the diagonal elements $\Bigl(V_{rr}(j)\Bigr)_{r=1}^d$ correspond to the auto-distance covariance function of $\{X_{t;r}\}$ and they express the dependence among the pairs $\bigl(X_{t;r}, X_{t+j;r}\bigr)$, $r = 1, \dots, d$.
  \item The off-diagonal elements $\Bigl(V_{rm}(0)\Bigr)_{r,m=1}^d$ measure the concurrent dependence between $\{X_{t;r}\}$ and $\{X_{t;m}\}$. If $V_{rm}(0) > 0$, $\{X_{t;r}\}$ and $\{X_{t;m}\}$ are concurrently dependent.
  \item For $j \in \mathbb{Z} - \{0\}$, $\Bigl(V_{rm}(j)\Bigr)_{r,m=1}^d$ measures the dependence between $\{X_{t;r}\}$ and  $\{X_{t+j;m}\}$. If $V_{rm}(j)=0$ for all $j \in \mathbb{Z} - \{0\}$, then $\{X_{t+j;m}\}$ does not depend on $\{X_{t;r}\}$.
  \item For all $j \in \mathbb{Z}$, $V_{rm}(j) = V_{mr}(j) = 0$ implies that $\{X_{t;r}\}$ and $\{X_{t+j;m}\}$ are independent. Moreover, for all $j \in \mathbb{Z}-\{0\}$, if $V_{rm}(j)=0$ and $V_{mr}(j)=0$ then $\{X_{t;r}\}$ and $\{X_{t;m}\}$ have no lead-lag relationship.
  \item If for all $j > 0$, $V_{rm}(j)=0$ but there exists some $j > 0$ such that $V_{mr}(j) > 0$, then $\{X_{t;m}\}$ does not depend on any past values of $\{X_{t;r}\}$, but $\{X_{t;r}\}$ depends on some past values of $\{X_{t;m}\}$.
\end{itemize}

%
%
%

As a closing remark, \citet{Davisetal} applied distance covariance methodology to stationary univariate and multivariate time series to study serial dependence under various choices of the weight function $\omega(\cdot,\cdot)$. One of their main results was to show that the asymptotic distribution of the empirical ADCV when applied to the residuals of a fitted  autoregressive process converges to a random variable whose distribution differs from that of ordinary ADCV.

\section{Test Statistics for Pairwise Dependence in Time Series}
\subsection{Introduction}
There are many available tests for testing serial independence in time series literature using both time domain and frequency domain methodologies, see \citet{Dag}. For a univariate time series, the most well known testing procedures are mainly based on the serial correlation coefficient including those by \citet{BoxPierce}
\begin{equation*}
\mbox{BP} = n\sum_{j=1}^{p}{\hat{\rho}^2(j)},
\end{equation*}
and \citet{LjungBox}
\begin{equation*}
\mbox{LB} = n(n+2)\sum_{j=1}^{p}{(n-j)^{-1}\hat{\rho}^2(j)},
\end{equation*}
among others, where $\rho(j) = \mbox{Cor}(X_t,X_{t+j})$ for $j \in \mathbb{Z}$ and $\hat{\rho}(j)$ denotes its empirical analogue \citep[pg.29]{BrockwellDavis}. Most of these procedures were extended to the multivariate case; see \citet{Mahdi} for an overview and a newly proposed correlation-based test. The multivariate Ljung-Box test statistic \citet{Hosking},\citet{LiMcLeod}, \citet{Li(2004)}
\begin{equation*}
\label{eq:mLB}
\mbox{mLB} = n^2\sum_{j=1}^{p}{(n-j)^{-1}\mbox{trace}\{\widehat{\Gamma}'(j)\widehat{\Gamma}^{-1}(0)\widehat{\Gamma}(j)\widehat{\Gamma}^{-1}(0)\}},
\end{equation*}
is widely used for testing $\Gamma(1) = \dots = \Gamma(p) = 0$, where $\Gamma(j) = E\Bigl[(\textbf{X}_{t+j}-\boldsymbol\mu)(\textbf{X}_{t}-\boldsymbol\mu)'\Bigr]$ is the auto-covariance matrix of $\{\textbf{X}_t\}$ and $\widehat{\Gamma}(j)$ denotes the sample autocovariance matrix (see \citet[pg. 397]{BrockwellDavis}). Although these tests perform well under linear and Gaussian processes, they behave poorly against general types of nonlinear dependencies including those with zero autocorrelation (ARCH, bilinear, nonlinear moving average processes); see \citet{Romano} and \citet{Shao}. \citet{Dag} pointed out that a standard procedure for increasing the power of the correlation based tests in ARCH type models is to compute the correlation of the squared observations. However, this strategy leads to a loss of power compared to the ordinary correlation based tests. \citet{Robinson} gave a variety of test statistics for testing serial correlation under the presence of conditional heteroskedasticity, whereas \citet{Escanciano} further proposed an automatic Portmanteau test that allows for nonlinear dependencies and automatic selection of lag order $p$.
\subsection{Asymptotic tests based on ADCV}
\label{sec:testsADCV}
Under the null hypothesis of pairwise independence between $\textbf{X}_t$ and $\textbf{X}_{t+j}$, \citet{Zhou} obtained the asymptotic distribution of the sample ADCV at fixed lag $j$ and shows that it has identical limiting distribution as that obtained by \citet{Szekely} for the case of independent data (recall \eqref{eq:asympDistrVn}). Under the alternative hypothesis that $\textbf{X}_t$ and $\textbf{X}_{t+j}$ are dependent, $n\widehat{V}_\textbf{X}^2(j)$ tends to infinity. However, \citet{Zhou} studied the behavior of ADCV in multivariate time series at a fixed lag. \citet{FokianosPitsillou} relaxed this assumption by defining the ADCV in the framework of generalized spectral domain following \cites{HongB} methodology.
It can be shown that serial dependence in time series can be characterized by the generalized spectral density function. In fact, it is well known that any deviation of the generalized spectral density from uniformity is a strong evidence of serial dependence. \citet{HongB} confirmed these facts by showing that the standard spectral density approach becomes inappropriate for non-Gaussian and nonlinear processes with zero autocorrelation. To bypass such problems he proposes a generalized spectral density that is defined based on \eqref{eq:sigma}, that is
\begin{eqnarray*}
\label{eq:spectral}
f(\omega,u,v) & = & \displaystyle \frac{1}{2\pi} \sum_{j=-\infty}^{\infty}{\sigma_j(u,v)e^{-ij\omega}}, ~~~ \omega \in [-\pi,\pi],
\end{eqnarray*}
provided that $\mbox{sup}_{(u,v) \in \mathbb{R}^2}\sum_j{\abs{\sigma_{j}^{(r,m)}(u,v)}} < \infty$. This 'generalized' spectral density can be defined for both linear and nonlinear processes and captures all pairwise dependencies while preserving the nice properties of the standard spectrum. The null hypothesis of pairwise independence can be interpreted in terms of the generalized spectral density; i.e.\ testing whether
\[f(\omega,u,v)= f_0(\omega,u,v) = \frac{\sigma_0(u,v)}{2\pi}.\]
To test for serial dependence, \citet{HongB} compared the \cites{Parzen} kernel-type estimators of $\hat{f}(\omega,u,v)$ and $\hat{f}_0(\omega,u,v)$ by the $L_2$-norm. 
The final test statistic takes the form
\begin{eqnarray}
\label{eq:Tn2}
\mbox{H}_{99} & = & \int_{\mathbb{R}^2}{\sum_{j=1}^{n-1}{(n-j)k^2(j/p)\abs{\hat{\sigma}_j(u,v)}^2} \omega(u,v) du dv},
\end{eqnarray}
where $p$ is a bandwidth or lag order and $k(\cdot)$ is a Lipschitz-continuous kernel function with the following properties:
$k: \mathbb{R} \rightarrow [-1,1]$ is symmetric and is continuous at 0 and all but a finite number of points, with $k(0)=1$, $\int_{-\infty}^{\infty}{k^2(z)dz} < \infty$  and $\mid k(z) \mid \leq C \mid z \mid^{-b}$ for large $z$ and $b > 1/2$.
\par
Distance measures obtained by characteristics functions have been also considered by \citet{Pinkse} and \citet{HongB}. \citet[pg.1206]{HongB} explained how tests based on the empirical characteristic function may have omnibus power against tests based on the empirical distribution function.
In addition to the test of serial dependence, \citet{HongB} proposed testing other aspects of dependence after substituting $\hat{\sigma}_j(u,v)$ in \eqref{eq:Tn2} by its derivatives $\hat{\sigma}_j^{(m,l)}(u,v)=\partial^{m+l}\hat{\sigma}_j(u,v)/\partial^mu\partial^lv$. Different choices of the indices $(m,l)$ and the weight function $\omega(u,v)=\omega_1(u)\omega_2(v)$ deliver a wide range of various hypotheses tests. For instance, choosing $(m,l)=(1,1)$ and both $\omega_1(u)$, $\omega_2(v)$ having the Dirac delta function, yields a test statistic for serial correlation.
To test for the martingale hypothesis which implies that the best predictor of future values of a time series,
in the sense of least mean squared error, is simply the current
value of the time series, put $(m,l) = (1,0)$ and assume that $\omega_1(u)$ has a Dirac density and $\omega_2(v)$ is an arbitrary increasing function that is integrable. For these latter choices of $\omega_1(\cdot)$ and $\omega_2(\cdot)$ and for the choices of $(m,l)=(3,0)$ and $(m,l)=(4,0)$, we can test for conditional symmetry and conditional heterokurtosis respectively; see \citet{HongB} for more details.

\par
Under the null hypothesis of independence, \citet{HongB} proved that a standardized version of $\mbox{H}_{99}$ given in \eqref{eq:Tn2} is asymptotically standard normally distributed, provided that $p=cn^{\lambda}$, for $c > 0$ and $\lambda \in (0,1)$. He further proved that the asymptotic power of the derived test statistic approaches unity; hence the test statistic is consistent. Recalling \eqref{eq:V}, one can observe that the statistic $\mbox{H}_{99}$ proposed by \citet{HongB} is similar to that of \citet{Szekely} with the main differences being the restriction to the univariate case and the choice of the weight function. Moreover, the number of lags used in the construction of $\mbox{H}_{99}$ is not fixed but increases with the sample size of the process. \citet{FokianosPitsillou} built a bridge between these two methodologies. They have chosen $\omega(\cdot,\cdot)$ of the form of \eqref{eq:W}, resulting in a test statistic based on the ADCV
\begin{eqnarray}
\label{eq:Tn}
\mbox{FP} & = & \sum_{j=1}^{n-1}{(n-j)k^2(j/p)\widehat{V}^2_X(j)}.
\end{eqnarray}
Clearly, the statistic $\mbox{FP}$ is a portmanteau type statistic, with each squared distance covariance being multiplied by a weight determined by a kernel function $k(\cdot)$. Under the null hypothesis that the process $X_t$ forms an i.i.d.\ sequence, a suitable standardization of $\mbox{FP}$ was proved to follow asymptotically a standard normal distribution, provided that $p=cn^{\lambda}$, for $c > 0$ and $\lambda \in (0,1)$. In addition, considering any type of pairwise dependence for the process, the authors showed that the proposed test statistic is asymptotically consistent. The latter test statistic is faster to compute than the one proposed by \citet{HongB} since the authors avoid a two dimensional integration. Although both test statistics, $\mbox{FP}$ and $\mbox{H}_{99}$ given in \eqref{eq:Tn} and \eqref{eq:Tn2} respectively, are defined for univariate $\alpha$-mixing processes, the proposed methodologies can be extended for multivariate processes as explained in the following. 

\par
Following a similar methodology for multivariate time series and considering the generalized spectral density matrix,
\citet{FokianosPitsillouB} proposed a test statistic based on the pairwise auto-distance covariance function
\begin{eqnarray*}
\label{eq:tildeTn}
\widetilde{\mbox{FP}} & = &  \sum_{r,m}{\sum_{j=1}^{n-1}{(n-j)k^2(j/p) \widehat{V}_{rm}^2(j)}} = \sum_{j=1}^{n-1}{(n-j)k^2(j/p)\mbox{trace}\{\widehat{V}^{*}(j)\widehat{V}(j)\}},
\end{eqnarray*}
where $\widehat{V}^{*}(\cdot)$ denotes the conjugate transpose of the sample distance covariance matrix $\widehat{V}(\cdot)$.
Under the null hypothesis of independence, a standardized version of $\widetilde{\mbox{FP}}$ follows asymptotically a standard normal distribution, provided that the numbers of lags employed in the construction of $\widetilde{\mbox{FP}}$ increases with the sample sizes, whereas under any type of alternative hypothesis the proposed test statistic has asymptotic power one.
\par
\citet{ChenHong} developed a nonparametric test for the Markov property of a multivariate time series based on the conditional characteristic function. Based on the fact that the correlation coefficient is suitable for Gaussian data but it fails for nonlinear cases, many authors defined the correlation function in a local sense, including \citet{TjostheimHufthammer}, \citet{BerentsenTjostheim}, \citet{Stove} and \citet{StoveTjostheim} among others. In the context of time series,
\citet{LacalDag} defined a new measure of dependence, the so-called local Gaussian autocorrelation that works well for nonlinear models. The authors compared the proposed test statistic to distance covariance function and found that they both work in a similar way.
\subsection{Other test statistics}
Many studies in the literature have considered the problem of measuring dependence between $X_t$ and $X_{t+j} \equiv Y_t$, in terms of the distance between the bivariate distribution of $(X_t,Y_t)$, $F_{X;Y}(x,y) = P(X_t \leq x, Y_t \leq y)$, and the product of their marginal distribution functions, $F_{X}(x) = P(X_t \leq x)$ and $F_{Y}(y) = P(Y_t \leq y)$. Well known distance measures for distribution functions are the Kolmogorov-Smirnov distance
  \begin{equation}
  \label{eq:KSdist}
  D_1(j) = \mbox{sup}_{(x,y) \in \mathbb{R}^2}\abs{F_{X;Y}(x,y) - F_X(x)F_Y(y)}
  \end{equation}
  and the Cramer-von Mises type distance
  \begin{equation}
  \label{eq:CvMdist}
  D_2(j) = \int_{\mathbb{R}^2}{\Biggl\{F_{X;Y}(x,y)-F_X(x)F_Y(y)\Biggr\}^2dF_{X;Y}(x,y)}.
  \end{equation}
Replacing the theoretical distribution functions with their empirical analogues
\begin{equation}
\label{eq:EDF}
\begin{aligned}
\widehat{F}_{X;Y}(x,y) & = \frac{1}{(n-j)}\sum_{t=1}^{n-j}{\mathbb{I}(X_t \leq x, Y_t \leq y)}, \\[1ex]
\widehat{F}_{X}(x) & = \frac{1}{(n-j)}\sum_{t=1}^{n-j}{\mathbb{I}(X_t \leq x)},
\end{aligned}
\end{equation}}
where $\mathbb{I}(\cdot)$ is the indicator function, we form the corresponding empirical estimates. Similar distance measures are obtained by employing density functions instead of distribution functions (see \citet{Skaug96} and \citet{Bagnato} for an overview).
\par
\citet{Skaug} extended the work by \citet{Blum} and considered the asymptotic behavior of the Cramer-von Mises type statistic \eqref{eq:CvMdist} at lag $j$, under ergodicity of $\{X_t\}$. Moreover, they constructed a test for pairwise independence among pairs $(X_t,X_{t+1})$, $(X_t,X_{t+2})$, \dots, $(X_t,X_{t+p})$ using the statistic
\begin{equation*}
\mbox{ST} = \sum_{j=1}^{p}{\widehat{D}_{2}(j)}
\end{equation*}
where $p$ is a fixed constant denoting the maximum lag order employed for the test. Under the null hypothesis that $\{X_t\}$ consists of i.i.d.\ random variables, $n \, \mbox{ST}$ converges in distribution (as $n \rightarrow \infty$) to an infinite mixture of chi-squared random variables. However, the latter test statistic is consistent against serial dependencies up to a finite order $p$. From a practical point of view, this may be restrictive since the actual serial dependence may be detected at lags larger than $p$. Motivated by the approach introduced by \citet{Skaug}, \citet{HongD} proposed a statistic where the number of lags increases with the sample size and different weights are given to different lags; in other words let
\begin{equation*}
\mbox{H}_{98}= \sum_{j=1}^{n-1}{(n-j)k^2(j/p)\widehat{D}_{2}^2(j)}
\end{equation*}
where $k(\cdot)$ is a kernel function with properties similar to the kernel function of \eqref{eq:Tn2}. He further proved that after proper standardization and for large $p$, provided that $p=cn^{\lambda}$, for $c > 0$ and $\lambda \in (0,1)$, the statistic is asymptotically standard normally distributed.
\par
In general, incorporating a large number of lags in the asymptotic theory is an issue nicely addressed by the frequency domain framework. 
\citet{Hong} proposed three test statistics for testing serial independence in univariate time series by considering a kernel-based standardized spectral density
\begin{equation*}
\label{eq:spectralEst}
\hat{f}(\omega) = \frac{1}{2\pi}\sum_{j=-(n-1)}^{(n-1)}{k(j/p)\hat{\rho}(j)\mbox{cos}(j\omega)} , \quad \omega \in [-\pi,\pi]
\end{equation*}
where the notation follows the notation of \eqref{eq:Tn2} and $\hat{\rho}(j)$ is the sample autocorrelation function of $\{X_t\}$ at lag $j$. Comparison of $\hat{f}(\omega)$ to $f_0(\omega) = 1/2\pi$ (spectral density of white noise process) via divergence measures yields test statistics that can be employed for testing that the process $\{X_t\}$ is white noise. In particular, he employs the quadratic norm, the Hellinger metric and the Kullback-Leibler information criterion, i.e.\
\begin{eqnarray*}
T_{1n} & = & \biggl[2\pi\int_{-\pi}^{\pi}{\Bigl\{\hat{f}(\omega)-f_0(\omega)\Bigr\}^2 d\omega}\biggr]^{1/2}, \\[1ex]
T_{2n} & = & \biggl[\int_{-\pi}^{\pi}{\Bigl\{\hat{f}^{1/2}(\omega)-f_0^{1/2}(\omega)\Bigr\}^2 d\omega}\biggr]^{1/2}, \\[1ex]
T_{3n} & = & -\int_{\omega:\hat{f}(\omega) > 0}{\mbox{ln}\Bigl(\hat{f}(\omega)/f_0(\omega)\Bigr)f_0(\omega) d \omega},
\end{eqnarray*}
respectively. A closed form expression for $T_{1n}$ is given by
\begin{equation*}
\label{eq:T1n}
\mbox{H}_{96} = n\sum_{j=1}^{n-1}{k^2(j/p)\hat{\rho}^2(j)}.
\end{equation*}
Compared to the classical portmanteau statistics, like BP and LB, all these suggested statistics, properly standardized, are asymptotically standard normally distributed and can be developed without imposing a specific alternative model. \citet{XiaoWu} proved that the asymptotic null distribution of $\mbox{H}_{96}$ in its standardized form, remains intact under the presence of serial correlation, whereas \citet{Shao} proved its robustness to conditional heteroscedasticity. Although \cites{Hong} tests incorporate an increasing number of lags, they are based on the classical correlation coefficient and are not consistent against all pairwise dependencies of unknown form. To avoid this problem, \citet{HongC} generalizes both the Cramer-von Mises and Kolmogorov-Smirnov type test statistics by employing the generalized spectral theory. Define a dependence measure analogous to \eqref{eq:sigma} by
\begin{equation}
\label{eq:rhoj}
\rho^{*}_j(x,y) = F_{X;Y}(x,y) - F_X(x)F_Y(y).
\end{equation}
Clearly, the distance dependence measure $\rho^{*}(\cdot,\cdot)$ defined in \eqref{eq:rhoj} vanishes only in the case where $X_t$ and $Y_t \equiv X_{t+j}$ are independent, leading to the observation that $\rho^{*}_j(\cdot,\cdot)$ can capture all pairwise dependencies. Replacing the theoretical distribution functions with the empirical ones given in \eqref{eq:EDF}, one can find the empirical analogue of \eqref{eq:rhoj}, $\hat{\rho}^{*}_j(\cdot,\cdot)$. Clearly, \cites{Skaug} approach is based on \eqref{eq:rhoj}.
\par
In a recent work, \citet{Dette} introduced a "new" spectrum as the Fourier transform of the so-called copula cross-covariance kernel
\begin{equation}
\label{eq:dette}
\rho_j^{U}(\tau_1,\tau_2) := \mbox{Cov}\Bigl(\mathbb{I}(U_t \leq \tau_1),\mathbb{I}(U_{t+j} \leq \tau_1)\Bigr),
\end{equation}
where $(\tau_1,\tau_2) \in (0,1)^2$ and $U_t := F(X_t)$. They proposed to estimate the corresponding spectral densities associated with measures \eqref{eq:rhoj} and \eqref{eq:dette} via Laplace periodogram \citep{Li(2008)}. In addition, they highlighted that in the case of using \eqref{eq:dette}, replacing the original observations with their ranks, potentially preserves the invariance property with respect to transformations of the marginal distributions.

\subsubsection{Multivariate Extension}

All univariate testing methodologies presented in this section can be extended to the multivariate case by considering the distances between $X_{t;r}$ and $Y_{t;m} \equiv X_{t+j;m}$ for $r,m=1,\dots,d$. The idea is to compare the pairwise bivariate distribution function of $(X_{t;r}, Y_{t;m})$, $F_{X_r;Y_m}(x,y) = P(X_{t;r} \leq x, Y_{t;m} \leq y)$, and the product of their marginal distribution functions, $F_{X_r}(x) = P(X_{t;r} \leq x)$ and $F_{Y_m}(y) = P(Y_{t;m} \leq y)$ via the dependence measures \eqref{eq:KSdist} and \eqref{eq:CvMdist}. The final test statistics can then be defined by summing over $r,m=1, \dots, d$. The multivariate extensions of $\mbox{ST}$ and $\mbox{H}_{98}$ are derived as

\begin{equation*}
\widetilde{\mbox{ST}} = \sum_{r,m=1}^{d}\sum_{j=1}^{p}{\widehat{D}_{2}^{(r,m)}(j)}
\end{equation*}
and
\begin{equation*}
\widetilde{\mbox{H}_{98}} = \sum_{r,m=1}^{d}\sum_{j=1}^{n-1}{(n-j)k^2(j/p)\Bigl(\widehat{D}_{2}^{(r,m)}(j)\Bigr)^2},
\end{equation*}

respectively, where $\widehat{D}_{2}^{(r,m)}(\cdot)$ is the Cramer-von Mises type distance given by
\begin{equation*}
  D_2^{(r,m)}(j) = \int_{\mathbb{R}^2}{\biggl\{F_{X_r;Y_m}(x,y)-F_{X_r}(x)F_{Y_m}(y)\biggr\}^2dF_{X_r;Y_m}(x,y)}.
  \end{equation*}

All other univariate test statistics mentioned in this section can be developed along the same lines. Although the development of the asymptotic theory of these test statistics is not in the scope of this review paper, we think that this is a challenging topic for further research.

\section{Applications}
\subsection{Univariate Case}
We investigate the finite sample performance of the statistics $\mbox{BP}$, $\mbox{LB}$, $\mbox{H}_{96}$, $\mbox{H}_{98}$, $\mbox{H}_{99}$, $\mbox{ST}$ and $\mbox{FP}$ presented in Section \ref{sec:testsADCV}. All simulations performed correspond to different sample sizes ($n=100, 500$). We employ ordinary bootstrap (with size $b=499$) to obtain critical values in order to study the size and the power of the statistics.
We first examine  the size  of the tests. Suppose that $\{X_t\}$ is an i.i.d.\ sequence of standard normal random variables. The spectral
density based test statistics, $\mbox{H}_{96}$, $\mbox{H}_{98}$, $\mbox{H}_{99}$ and $\mbox{FP}$, are all computed based on the Bartlett kernel with three different choices of bandwidth $p=[3n^\lambda]$, where $\lambda=0.1$, $0.2$ and $0.3$. Table \ref{tab:size} shows the achieved type I error rates at $5\%$ nominal level. All  test statistics achieve nominal sizes quite adequately especially for large bandwidth parameters.

\begin{table}[t]
\small
\centering
\caption{\small{Achieved type I error of the test statistics for testing the hypothesis that the data are i.i.d. The data are generated by the standard normal distribution. Significance level is set to $\alpha=0.05$. Achieved significance levels are given in percentages. The results are based on $b=499$ bootstrap replications and 1000 simulations.}}
\begin{tabular}{cccccccccc}
  \hline
  \hline
  $n$ & $p$ & & $\mbox{BP}$ & $\mbox{LB}$ & $\mbox{H}_{96}$ & $\mbox{H}_{98}$ & $\mbox{H}_{99}$ & $\mbox{ST}$ & $\mbox{FP}$ \\
  \hline
  100 & 5 & & 4.5 & 5.0 & 4.9 & 4.6 & 4.2 & 6.0 & 4.8 \\
   & 8 & & 4.9 & 5.2 & 5.1 & 5.8 & 4.3 & 5.9 & 4.7 \\
   & 12 & & 4.0 & 5.6 & 5.6 & 3.9 & 4.1 & 3.4 & 4.0 \\[2ex]
  500 & 6 & & 3.5 & 5.4 & 5.6 & 5.8 & 4.5 & 4.8 & 4.3 \\
   & 11 & & 4.1 & 4.4 & 5.0 & 5.2 & 5.6 & 6.1 & 5.0 \\
   & 20 & & 3.3 & 5.5 & 4.1 & 5.1 & 5.7 & 5.1 & 4.8 \\
  \hline
\end{tabular}
\label{tab:size}
\end{table}

\par
For investigating the power of the tests we consider the NMA(2) model given in \eqref{eq:NMA2model} and the following additional models:
\begin{itemize}
  \item AR(1) -model
 \begin{equation}
 \label{eq:AR1model}
 X_t = 0.4X_{t-1} + \epsilon_t,
 \end{equation}
  \item ARCH(2)-model
 \begin{equation}
 \label{eq:ARCH2model}
 X_t = \sigma_t \epsilon_t, \quad \sigma_t^2 = 0.5 + 0.8X_{t-1}^2 + 0.1X_{t-2}^2
 \end{equation}
\end{itemize}
where $\{\epsilon_t\}$ is a sequence of i.i.d.\ standard normal random variables. Note that \eqref{eq:AR1model} corresponds to a first-order autoregressive process, \eqref{eq:ARCH2model} corresponds to an autoregressive conditional heteroscedastic model of order two (see \citet{Engle}). 
It is well known that the process $\{X_t\}$ generated by \eqref{eq:NMA2model} consists of a sequence of dependent but uncorrelated random variables.
Table \ref{tab:power}
show the power of all test statistics considered for various sample sizes and bandwidth parameters by
models \eqref{eq:NMA2model}, \eqref{eq:AR1model} and \eqref{eq:ARCH2model} respectively. It is clear that when the data are generated by the nonlinear models NMA(2) and ARCH(2) the test statistics $\mbox{FP}$ and $\mbox{H}_{99}$ perform better, whereas in the case where the data are generated by the linear AR(1) model all test statistics exhibit similar power.

\begin{table}[h]
\small
\centering
\caption{Empirical power ($\%$) of all test statistics of size  $5\%$. The  bandwidth $p$ is  $[3n^\lambda]$, $\lambda=0.1$, $0.2$ and $0.3$.
The results are based on $B=499$ bootstrap replications for each of  $1000$ simulations. The test statistics $\mbox{H}_{96}$, $\mbox{H}_{99}$, $\mbox{FP}$ and $\mbox{ST}$ are calculated  by employing
the Bartlett kernel.}{
\begin{tabular}{@{}rccccccccc@{}} \hline \hline
$n$ & \multicolumn{3}{c}{100} &  \multicolumn{3}{c}{200} & \multicolumn{3}{c}{500}  \\
$p$ & 5  & 8  & 12  &   6  &  9  & 15  & 6 & 11 &20 \\[2pt]
& \multicolumn{3}{c}{} &  \multicolumn{3}{c}{Model \eqref{eq:NMA2model}} & \multicolumn{3}{c}{} \\
$\mbox{BP}$   & 23.5   &  20.5   & 17.8 &   27.9 &   22.7 &  20.5  & 35.8 &  33.3  & 25.1\\
$\mbox{LB}$  & 22.8   &  21.8   & 16.9 &   27.9 &   25.7 &  21.2  &  35.4 & 30.2 & 24.6\\
$\mbox{H}_{96}$  & 36.9   &  34.7   & 31.4 &   42.3 &   40 &  36.2  &  47.5 & 42.6 & 37.6\\
$\mbox{H}_{98}$  & 47.4   &  43.1   & 37.7 &   93.9 &   85.8 &  75.5  &  100 & 100 & 100\\
$\mbox{H}_{99}$  & 97   &  82   & 59 &   100 &   100 &  100  &  100 & 100 & 100\\
$\mbox{ST}$  & 26.1   &  20   & 17.5 &   53.7 &   43.9 &  31.2  &  100 & 98.3 & 84.4\\
$\mbox{FP}$  & 89.2   &  72.8   & 49.5 &   99.8 &   99.6 &  95  &  100 & 100 & 100\\[3pt]
& \multicolumn{3}{c}{} &  \multicolumn{3}{c}{Model \eqref{eq:AR1model}} & \multicolumn{3}{c}{} \\
$\mbox{BP}$   & 88.3  & 81.2  & 71.6  & 99.6  & 99.6  & 98.1  & 100   & 100  & 100  \\
$\mbox{LB}$  & 87.4   & 84.2 & 72 & 99.3 & 98.8 & 98 & 100  & 100 & 100\\
$\mbox{H}_{96}$  & 94.4   & 94.2 & 93.2 & 99.9 & 100 & 100 & 100  & 100 & 100\\
$\mbox{H}_{98}$  & 95   & 93.3 & 92.5 & 100 & 100 & 100 & 100  & 99.8 & 100\\
$\mbox{H}_{99}$  & 81.5   & 75.7 & 68.8 & 98.6 & 98 & 96.1 & 100  & 100 & 100\\
$\mbox{ST}$  & 80.5   & 76.1 & 67.1 & 98.2 & 97.2 & 93.3 & 100  & 100 & 100\\
$\mbox{FP}$  & 93.5   & 87.7 & 84.8 & 100 & 99.8 & 99.4 & 100  & 100 & 100\\[3pt]
& \multicolumn{3}{c}{} &  \multicolumn{3}{c}{Model \eqref{eq:ARCH2model}} & \multicolumn{3}{c}{} \\
$\mbox{BP}$   & 23.4 &  19.4 &  17.3 & 30.1  & 24.8    & 23.5  & 40.6   & 34.2  & 27.1\\
$\mbox{LB}$  & 21.4 &  18.7 &  15.9 & 28.9  & 24.6  & 20.7  & 38.2   &35.9 & 26.4\\
$\mbox{H}_{96}$  & 37.1 &  35.2 &  35.6 & 48  & 48.2  & 47.9  & 64.7   &63.1 & 63.6\\
$\mbox{H}_{98}$  & 24.9 &  22.9 &  19.3 & 40.1  & 39.7  & 35.4  & 89   &83.8 & 72.2\\
$\mbox{H}_{99}$  & 71 &  74 &  60 & 99  & 96  & 86  & 100   &100 & 100\\
$\mbox{ST}$  & 20.8 &  14.3 &  14.3 & 29.1  & 27.3  & 20.8  & 67.3   &54.5 & 39\\
$\mbox{FP}$  & 59.8 &  54 &  45.3 & 89  & 81.3  & 76  & 99.9   &100 & 100\\
\end{tabular}}
\label{tab:power}
\end{table}

\vspace{-0.5cm}
\subsection{A Data Example}
\begin{figure}
  \centering
  \includegraphics[height=7cm,width=\linewidth]{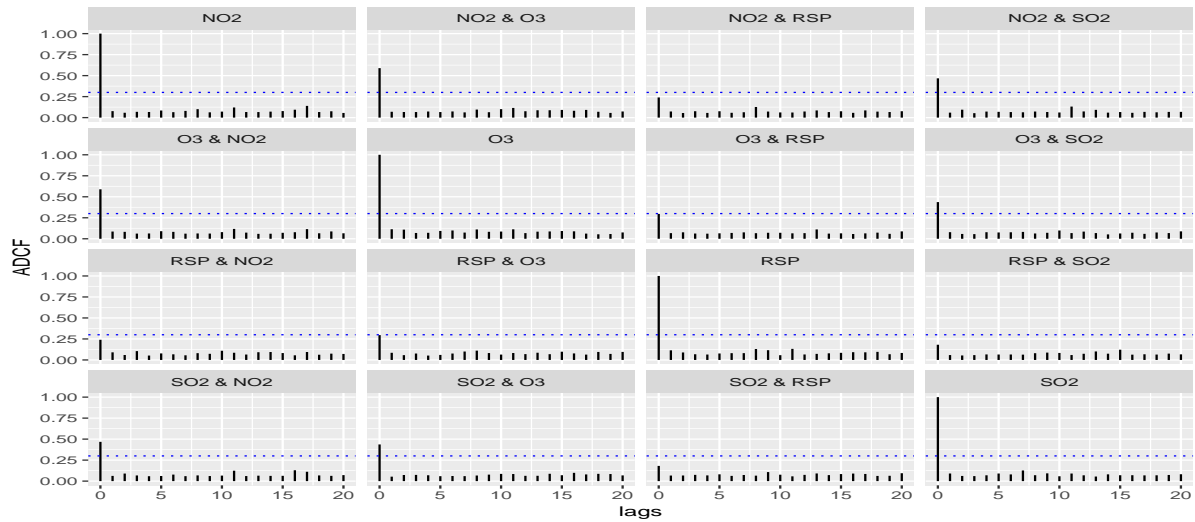}
  \caption{The ADCF plots of the residuals of the VAR(7) model for the four-dimensional air pollution time series data.}
  \label{fig:residualsPlot}
\end{figure}

We analyze  air pollution data on four pollutants, the sulfur dioxide ($SO_2$), ozone ($O_3$), nitrogen dioxide ($NO_2$) and respirable suspended particulates ($RSP$) measured daily from the monitoring station at Tsuen Wan in Hong Kong from February 2015 to December 2016. The time series data can be obtained from the Environmental Protection Department of Hong Kong (\url{http://epic.epd.gov.hk/EPICDI/air/station/?lang=en}). Each time series consists of 700 observations.
We  use a log-transformation to the data and applied a first-order differencing on the series to remove seasonal patterns.
Assuming that the four dimensional series follow a vector autoregressive model (VAR) and employing the AIC to choose its order, we obtain that a seventh order model would be appropriate for the data. After a VAR(7) model fit, we examine the ADCF plots of the residuals (Figure \ref{fig:residualsPlot}). The plot does not show any evidence of serial dependence. Cnstructing multivariate tests of independence among the residuals, all test statistics used, $\widetilde{\mbox{FP}}$, $\mbox{mLB}$, $\widetilde{\mbox{ST}}$ and $\widetilde{\mbox{H}_{98}}$, yield large $p$-values (Table \ref{tab:pvalAIR}) suggesting the sufficiency of this model fit. Some of the methods are implemented in the R package \verb"dCovTS"--see
\citet{dCovTS}.

%

\begin{table}
\small
\centering
\caption{\small{$P$-values of tests of independence among the residuals after fitting a VAR(7) model to the four dimensional time series. All results are based on $B=499$ bootstrap replications. Both statistics $\widetilde{FP}$ and $\widetilde{H}_{98}$ are calculated based on the Bartlett kernel.}}
\begin{tabular}{ccccc}
  \hline \hline
  $p$ & $\widetilde{\mbox{FP}}$ & $\mbox{mLB}$ & $\widetilde{\mbox{ST}}$ & $\widetilde{\mbox{H}_{98}}$ \\
  \hline
  6 & 0.332 & 0.999 & 0.528 & 0.584 \\
  11 & 0.368 & 0.973 & 0.512 & 0.546 \\
  22 & 0.104 & 0.576 & 0.554 & 0.128 \\
  \hline
\end{tabular}
\label{tab:pvalAIR}
\end{table}

\bibliographystyle{chicago}
\bibliography{refs_abb}

\end{document}